\documentclass[fleqn,10pt]{wlscirep}
\usepackage[utf8]{inputenc}
\usepackage[T1]{fontenc}
\usepackage{subcaption}
\usepackage[numbers]{natbib}
\usepackage{amsmath}
\usepackage{bbm}

\newcommand{\LanB}{\textit{LanB1}\textsuperscript{\textit{Df}}}

\title{From dispersal to patrol: the extracellular matrix shapes an annealing-like search by \textit{Drosophila} embryonic macrophages}

\author[1,*]{Nickolay Korabel}
\author[2,**]{Tom H. Millard}
\author[3,4,***]{Thomas Andrew Waigh}
\affil[1]{Department of Physics, Astronomy and Mathematics, University of Hertfordshire, Hatfield, UK}
\affil[2]{Faculty of Biology, Medicine and Health, The University of Manchester, Manchester, UK}
\affil[3]{Biological Physics, Department of Physics and Astronomy, The University of Manchester, Manchester, UK}
\affil[4]{Photon Science Institute, The University of Manchester, Manchester, UK}
\affil[*]{m.korabel@herts.ac.uk}
\affil[**]{Tom.Millard@manchester.ac.uk}
\affil[***]{t.a.waigh@manchester.ac.uk}

\begin{abstract}
Hemocytes, the macrophages of \textit{Drosophila}, disperse through the embryo
and then patrol its tissues, moving along basement membranes that they help to
build. Because the persistence ($H$) and amplitude ($D$) of hemocyte motion
vary along each track, it is difficult to describe how movement changes during
development and how it depends on basement membrane composition. We analysed
trajectories, obtained by selective plane illumination microscopy, of hemocytes
in control embryos and in embryos deficient in the basement membrane components
laminin (\textit{LanB1}-deficient) or collagen~IV (collagen-deficient). A
physics-informed neural network estimates $H$ and $D$ in short windows, each
with a calibrated uncertainty. From them we construct an effective diffusivity
$D_{\mathrm{eff}}$, defined for any anomalous diffusion as the diffusivity of
the unpredictable part of the motion. Its error is uncorrelated with that of
$H$ and acts as an effective temperature. Cells decelerate during
development by cooling (falling $D_{\mathrm{eff}}$) and forgetting (falling
$H$). Control cells cool gradually while staying persistent.
\textit{LanB1}-deficient cells cool further, forget, lose the coupling between
speed and persistence and become nearly Brownian. Collagen-deficient cells cool
but keep most of their persistence. Hemocyte movement thus resembles an
annealed search, in which laminin, rather than collagen~IV, keeps patrolling
cells persistent.

\end{abstract}

\begin{document}

\flushbottom
\maketitle
%
%
\thispagestyle{empty}


\section*{Introduction}

Hemocytes are the macrophages of \textit{Drosophila melanogaster}. In the
embryo they appear in the head mesoderm, disperse along stereotyped routes
until they populate the whole embryo, and then remain in the tissues, where
they clear apoptotic corpses, respond to wounds and deposit extracellular
matrix~\cite{wood2007drosophila,stramer2005live,evans2014drosophila}. The
embryo thus provides a genetically tractable system in which a population of
immune cells switches, within a few hours, from colonising a tissue to
surveying it. The cues that guide the dispersal phase have been studied
extensively~\cite{wood2007drosophila,evans2014drosophila}. Less is known
about how the motility of individual cells is reorganised across the
transition, and whether that reorganisation follows a search strategy that
can be described in physical terms. The extracellular matrix along which
hemocytes migrate is the basement membrane, a thin sheet built mainly from
laminin, type~IV collagen, perlecan and nidogen. Hemocytes are a major source
of it: as they disperse they deposit most of the embryo's collagen~IV and
about half of its perlecan~\cite{matsubayashi2017moving}, and the laminin
they secrete strengthens their own migration~\cite{sanchez2017drosophila}.
In \textit{Drosophila}, LanB1 is the only laminin $\beta$ subunit, so embryos deficient in it lack functional laminin. Because the cells both build this matrix and move along it, the basement membrane is a natural candidate for shaping the change in motility, and embryos deficient in LanB1 or collagen~IV allow this to be tested.

Asking how the matrix shapes motility requires a quantitative description of
the search itself, and physics offers a vocabulary for search strategies ~\cite{klages2008anomalous}.
\textit{Escherichia coli} alternates between straight runs and random
reorientations~\cite{berg1972chemotaxis,Waigh_2024},
effector T cells perform generalized L\'evy walks~\cite{harris2012generalized}, and
intermittent strategies that alternate fast relocation with slow local
exploration can minimise the time needed to find sparse
targets~\cite{benichou2011intermittent,viswanathan1999optimizing}. A strategy of a different kind
is familiar from optimisation problems in computer science. In \emph{simulated
annealing}, a searcher moves with a noise amplitude set by a temperature that
is lowered over time~\cite{kirkpatrick1983optimization,geman1984stochastic}. At high temperature it
crosses barriers freely and explores the whole energy landscape; as the
temperature falls, it becomes sensitive to progressively shallower features
and settles where the landscape is favourable. Applied to hemocytes, this
picture predicts an early phase of vigorous, weakly constrained motion that
disperses the cells, followed by a gradual cooling that allows them to settle
and patrol. Testing the prediction requires two quantities for every cell at
every stage of development: a temperature-like measure of how strongly the
cell moves, and a measure of how much the cell remembers its previous
direction.

In studies of cell migration these two quantities are usually speed and
persistence, and both are measured using the mean squared displacement (MSD). The
MSD of a trajectory grows as $\langle r^{2}(t)\rangle\propto t^{\alpha}$, where
the anomalous exponent $\alpha=1$ corresponds to a random walk and $\alpha=2$
to ballistic (straight-line) motion ~\cite{waigh2014physics}. In the persistent random walk model the MSD crosses
over from ballistic to diffusive scaling at a single persistence
time~\cite{selmeczi2005cell}, and the local slope $\alpha(t)$ is therefore used to
follow persistence across time scales. In \textit{Dictyostelium},
$\alpha\approx1.5$ at short times in all conditions, but only cells moving
towards a source of directional cues kept $\alpha\approx1.6$ beyond a few
minutes; without a cue $\alpha$ decayed towards~1~\cite{mccann2010cell}. Many cell
types, however, remain superdiffusive over long times, with power-law velocity
correlations that a single persistence time cannot
capture~\cite{dieterich2008anomalous,wu2014three}. Speed and persistence are also not
independent. Across cell types, faster cells turn less~\cite{maiuri2015actin}, so
the two are correlated on the same trajectories~\cite{gorelik2014quantitative}.

Fractional Brownian motion (FBM) makes this description precise. The mean squared
displacement is $\langle r^2(t)\rangle = 2dDt^{2H}$, so that $\alpha=2H$ and \emph{d} is the number of dimensions. The
Hurst exponent $H$ measures directional memory, with persistent motion for
$H>1/2$, antipersistent motion for $H<1/2$ and Brownian motion at $H=1/2$,
and the generalized diffusion coefficient $D$ sets the amplitude of the
motion~\cite{metzler2004restaurant,metzler2014anomalous,waigh2023heterogeneous}.
Unlike the persistent random walk, FBM motion has no single
persistence time. Its memory extends over all time scales, as do the
correlations observed in migrating cells. In earlier work we imaged
hemocytes by selective plane illumination microscopy in control embryos and
in two mutants: \textit{LanB1}-deficient, which removes a laminin subunit and
disrupts the extracellular matrix, and \textit{SCAR}$^{\Delta37}$, which
impairs actin-driven motility~\cite{korabel2022hemocytes}. The motion was well
described by FBM that is heterogeneous in both time
and space. The distributions of displacements were non-Gaussian and were
reproduced by FBM with a Weibull distribution of
generalized diffusion coefficients. Local exponents, estimated by a neural
network in windows sliding along each track~\cite{han2020deciphering,korabel2025deep},
oscillated in time, with alternating periods of more or less persistent
motion. On average, the motion became less persistent as development
proceeded. Hemocytes that touched underwent synchronised contact inhibition
of locomotion, and both mutations disrupted the collective motion and
reduced its persistence. These results established that hemocyte motility
has to be followed locally, cell by cell and window by window, for a detailed
understanding of their behaviour. They also left open the question this work
addresses: as the cells lose persistence, do they also slow down?

Two obstacles prevented an answer. First, a window of twenty frames carries
an uncertainty comparable to the biological variation one wishes to resolve,
and the earlier local estimates carried no error bar. A change along a track
therefore could not be distinguished from a fluctuation of the
estimator~\cite{seckler2022bayesian}. Second, estimated $H$ and $D$ are correlated. 
$D$ carries units of $\mathrm{L}^2\,\mathrm{T}^{-2 H}$ ($L$ is length and $T$ is time), which differ whenever $H$ differs, so a cell that loses persistence also changes the units in which its diffusion coefficient is expressed. In a
short window, moreover, an error in $H$ is compensated almost entirely by an
error in $D$. The resulting correlation between $H$ and $\log D$ is a
property of the measurement, not of the cells, and even its sign depends on
the frame rate. It must not be confused with the biological coupling between
speed and persistence~\cite{maiuri2015actin}, which can only be measured once the
two have been separated. A change in $D$ during development therefore cannot
be interpreted until it has been separated from the change in $H$ that
accompanies it.

Here we overcome both limitations and follow hemocyte motility across six
hours of embryonic development, far longer than in our earlier
study~\cite{korabel2022hemocytes}. A physics-informed neural network estimates
$H$, $D$ and the localization error from short three-dimensional windows of
twenty frames, and returns with each estimate an uncertainty calibrated on
simulated trajectories. The network is trained on two terms: a supervised
loss on simulated parameters, and the exact likelihood of fractional Gaussian
noise, which ties every estimate to the observed steps themselves and needs
no ground truth. From its output we construct an effective diffusivity
$D_{\mathrm{eff}}$. Its definition is not specific to FBM. It applies to any anomalous diffusion process with finite second moments,
with stationary increments, as in FBM, or non-stationary ones, as in
continuous-time random walks and scaled Brownian motion. For FBM it takes a
closed form in $H$ and $D$, which we use throughout. The units of $D$
($\mu\mathrm{m}^2\,\mathrm{s}^{-2 H}$) change with $H$. $D_{\mathrm{eff}}$ is
always measured in $\mu\mathrm{m}^2\,\mathrm{s}^{-1}$, reduces to the ordinary diffusion coefficient for Brownian motion ($H=1/2$), and its estimation error
is, to leading order, uncorrelated with that of $H$. $D_{\mathrm{eff}}$ also has a direct physical meaning that holds for every process. Specifically, $2\Delta t\,D_{\mathrm{eff}}$ is the average variance of the part of each step that cannot be predicted from
the preceding steps. $D_{\mathrm{eff}}$ therefore measures how much fresh
randomness a cell injects per unit time, and $H$ separately records how
strongly each step remembers the past. By analogy with the Einstein relation,
$D_{\mathrm{eff}}$ defines an effective temperature
$T_{\mathrm{eff}}=\gamma D_{\mathrm{eff}}/k_B$, where $\gamma$ is a friction
coefficient and $k_B$ is Boltzmann's constant. The network is described in
Methods and Supplementary Sec.~S1, and the general definition of
$D_{\mathrm{eff}}$, its FBM form and its properties in Supplementary Sec.~S2.

With these tools, the transition from dispersal to patrol appears as a
trajectory through the plane of temperature ($T_{\mathrm{eff}}$) and memory
($H$). Early hemocytes are hot and persistent. As development proceeds,
$D_{\mathrm{eff}}$ falls in both the control and the \textit{LanB1}-deficient embryos, but the two genotypes cool along different paths. Control cells
retain their directional memory, whereas \textit{LanB1}-deficient cells cool further and become nearly Brownian. This refines the earlier observation that
hemocytes lose persistence during development~\cite{korabel2022hemocytes}. The
loss of memory is accompanied by cooling, the two can now be measured
separately, and their relative contributions differ between genotypes. We
interpret the decreasing effective temperature as an \emph{annealed search},
in which vigorous early motion disperses the population and progressive
cooling lets the cells settle into patrolling their local environments. We
also discuss an alternative explanation, involving an increase in friction as
the tissue matures, and identify measurements that can distinguish the two.

\section*{Results}
\label{sec:results}

\subsection*{Effective temperature and memory can be estimated from twenty-frame windows}

Anomalous diffusion is characterised by the mean squared displacement in $d$ dimensions
\begin{equation}  \langle\mathbf r^2(t)\rangle=2dDt^{\alpha},
  \label{eq:msd}
\end{equation}
where the
anomalous exponent $\alpha$ describes the memory of the motion and the
generalized diffusion coefficient $D$ its amplitude. For fractional Brownian motion (FBM) $\alpha=2H$, with $H$ the
Hurst exponent. Following how these two quantities change along a track raises
two problems. First, $D$ has units of $\mathrm L^2\,\mathrm T^{-\alpha}$, which
differ whenever $\alpha$ differs, so values of $D$ from cells or windows with
different $\alpha$ cannot be compared. Expressing time in a unit $c$ times
larger, shifts $\log D$ by $\alpha\log c$ ($\log_{10}$ is used
throughout), a different amount for every
$\alpha$. Second, the estimates of $\alpha$ and $D$ are not
independent, because both come from the same displacements. A short track
shows clearly the size of a typical step, but much less clearly how fast
displacements grow with lag time, and that growth rate is what determines
$\alpha$. An error in $\alpha$ therefore tilts the fitted mean squared
displacement about a pivot lag $\tau^*$, and $\log D$ shifts to compensate,
$\delta\log D\simeq-\log\tau^*\,\delta\alpha$ (see Supplementary Information). Longer
windows decrease both errors but leave the coupling intact. Correlations of
this kind between estimated exponents and diffusion coefficients,
$D\propto\exp(c_1\alpha+c_2)$, have been reported in several
single-particle-tracking experiments. They arise generically when both
parameters are estimated from the same finite
trajectory~\cite{lanoiselee2025super,han2026maximum}, and can be strong enough
to make a negative correlation between the true parameters appear
positive~\cite{lanoiselee2025super}. What is needed is a measure of the
amplitude of motion that has the same units for every trajectory and whose
estimation error is not inherited from that of $\alpha$.

We therefore introduce an effective diffusivity defined for any motion with
finite second moments, whether its increments are stationary, as for
fractional Brownian motion (FBM), or not, as for continuous-time random walks (CTRW) and L\'evy walks. For $N$ consecutive increments of one coordinate, sampled at interval
$\Delta t$, with covariance matrix $\Sigma$,
\begin{equation}  D_{\mathrm{eff}}=\frac{(\det\Sigma)^{1/N}}{2\Delta t}.
  \label{eq:Deff_gen_intro}
\end{equation}
For normal diffusion the increments are uncorrelated, so
$\Sigma=2D\Delta t\,I_N$, with $I_N$ the $N\times N$ identity, so
$(\det\Sigma)^{1/N}=2D\Delta t$ is simply the variance of a single step, and
$D_{\mathrm{eff}}=D$. When the increments are correlated, part of each step can
be predicted from the preceding ones. Because $\det\Sigma$ is the product of
the variances of the parts of successive steps that cannot be predicted from
the preceding ones, $(\det\Sigma)^{1/N}$ is then smaller than the step variance,
and $D_{\mathrm{eff}}$ is the diffusivity of the unpredictable part of the
motion. It has units of $\mathrm L^2\,\mathrm T^{-1}$ for every process.
Moreover, whenever the covariance is an amplitude multiplied by a matrix that
depends only on the memory parameters, $\log D_{\mathrm{eff}}$ is the measure of
amplitude whose estimation error is, to leading order, uncorrelated with the
errors of those parameters (Supplementary Sec.~S2).

For FBM the covariance has this form, $\Sigma=2D\Delta t^{2H}R(H)$, where
$R(H)$ is the correlation matrix of fractional Gaussian noise, and
Eq.~\eqref{eq:Deff_gen_intro} takes the form
\begin{equation}
  D_{\mathrm{eff}}=D\,\Delta t^{\,2H-1}\bigl[\det R(H)\bigr]^{1/N}.
  \label{eq:Deff}
\end{equation}
$D_{\mathrm{eff}}$ equals $D$ at $H=1/2$, has units of
$\mu\mathrm m^2\,\mathrm s^{-1}$ at every $H$. Importantly, estimation error is, to
leading order, uncorrelated with that of $H$. The $H$-dependent correction
$\Delta t^{2H-1}[\det R(H)]^{1/N}$ is fixed by the pivot lag $\tau^*$, so the
transformation removes exactly the coupling described above (Supplementary
Secs.~S1). The factor $[\det R(H)]^{1/N}$ is the average fraction of
a step that cannot be predicted from the preceding steps. 

\begin{figure}[t!]
  \centering
  \includegraphics[width=\textwidth]{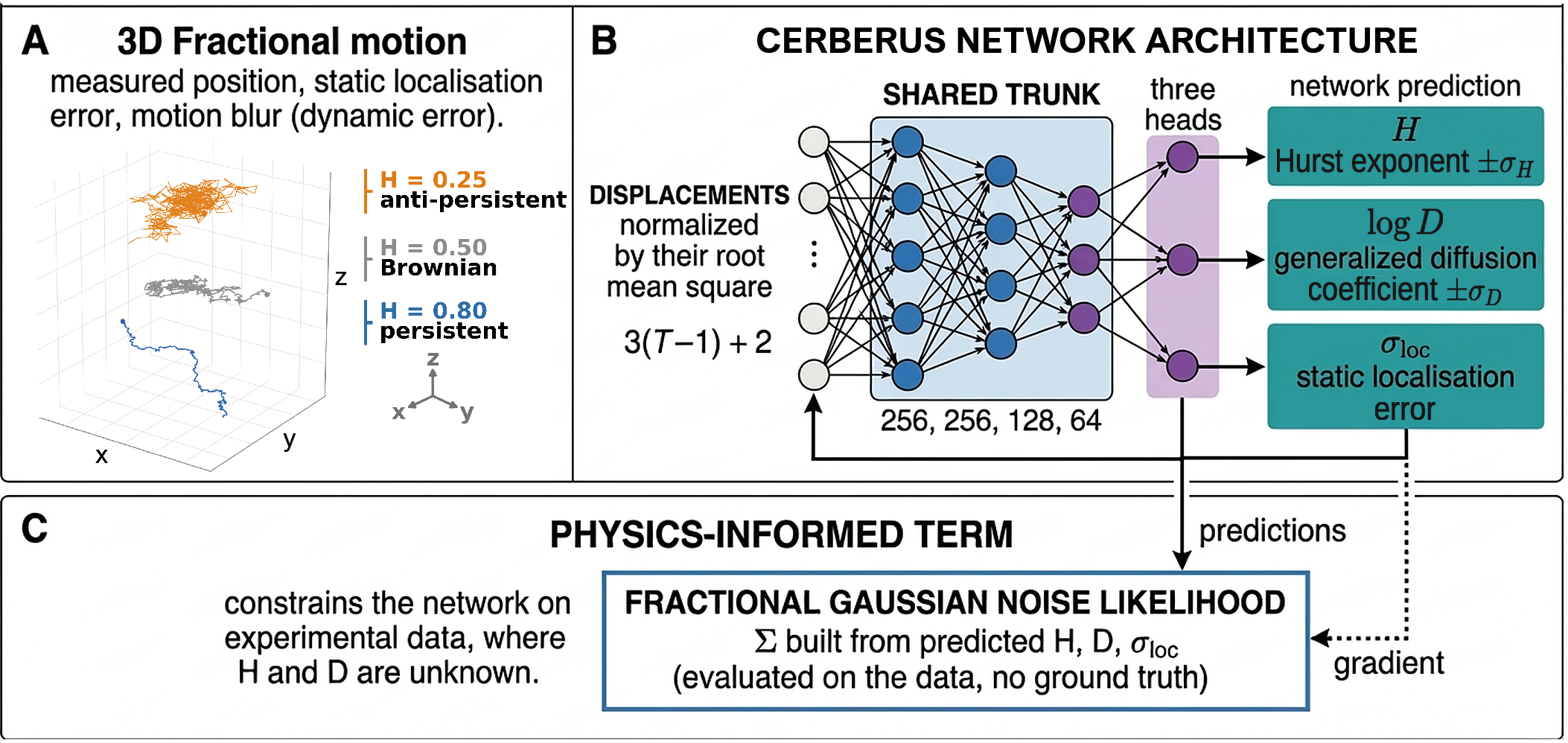}
  \caption{\textbf{A physics-informed neural network for the simultaneous
  estimation of the Hurst exponent, the generalized diffusion coefficient and
  the localization error.}
  \textbf{(A)} Three-dimensional FBM trajectories simulated with $H = 0.25$, $0.50$ and
  $0.80$, corresponding to anti-persistent, Brownian and persistent motion respectively.
  Open circles mark the start of each trajectory and filled circles the end.
  The measured positions include a static localization error.
  \textbf{(B)} Schematic of the Cerberus neural network. The input layer is loaded with the $3(T-1)$
  displacements of a window of $T$ points, normalized by their root mean
  square, together with two scalars which state whether the localization error
  is known and, if so, its size relative to the root mean square. A shared
  trunk of four fully connected layers with $256$, $256$, $128$ and $64$ units
  feeds three heads, which predict $H$, the logarithm of the generalized diffusion coefficient
  $D$ and the localization error $\sigma_{\mathrm{loc}}$. The numbers of nodes
  drawn are illustrative. The first two heads also predict a standard
  deviation, so every estimate carries its own uncertainty.
  \textbf{(C)} The physics-informed term. The predicted $H$, $D$ and
  $\sigma_{\mathrm{loc}}$ are used to build the covariance matrix $\Sigma$ of
  the observed displacements, and the exact fractional Gaussian noise
  likelihood is evaluated on the data themselves. Since this term does not use
  the ground truth, it constrains the network on experimental trajectories,
  where $H$ and $D$ are unknown, as well as on the simulated trajectories used
  for training.}
  \label{fig:fig0}
\end{figure}

To estimate $H$ and $D$ for short trajectories, we used a physics-informed neural
network (PINN) with a Cerberus architecture (Fig.~\ref{fig:fig0}). For each window it returns $H$,
$\log D$ and the static localization error $\sigma_{\mathrm{loc}}$ (due, for
example, to the finite resolution of the microscope), together with the
standard deviations $\sigma_{H}$ and $\sigma_{\log D}$.
The network sees only displacements normalized by their own root mean square,
so a single trained network can be applied at any frame rate and over decades
of $D$. The uncertainties are trained with a heteroscedastic loss and are
reported without \emph{post-hoc} rescaling (Supplementary Sec.~S1). On
simulated three-dimensional FBM trajectories with 20-point windows and a
signal-to-noise ratio of 20, the network recovers $H$ with a mean absolute
error of $0.065$ and $\log D$ with $0.23$\,dex, and is essentially unbiased
except very near the edges of the $H$ range. 
The stated uncertainties are calibrated, and the observed error tracks the
stated uncertainty across the whole ranges of $H$ and $D$ (Supplementary
Sec.~S1). The localization error, by contrast, is only weakly identifiable from
twenty frames, and the network estimate stays close to its prior. We therefore
report it but do not interpret it in detail.
\begin{figure}[t!]
    \centering    \includegraphics[width=\textwidth]{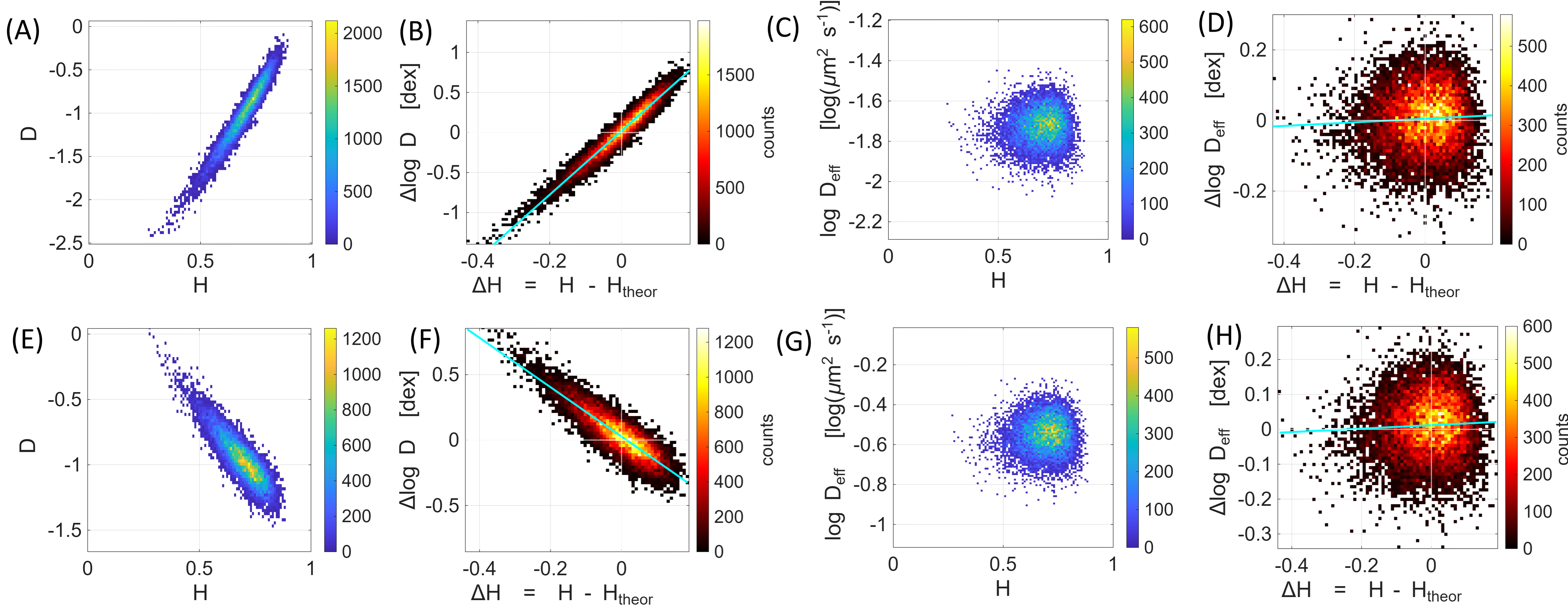}
\caption{\textbf{Joint probability distributions and estimation errors of fractional Brownian motion (FBM) parameters for different sampling time intervals.}
The theoretical parameters were
$H_{\mathrm{theor}}=0.7$ and
$\log D_{\mathrm{theor}}=-1$.
\textbf{(A,E)} Joint probability distributions of the predicted Hurst exponent
$H$ and predicted logarithm of the diffusion coefficient
$\log D$ for trajectories generated with
$dt=0.02$ and $dt=17$, respectively. The units of $D$ and $D_{theor}$ are
($\mu\mathrm{m}^2\,\mathrm{s}^{-2H}$) and change with $H$.
\textbf{(B,F)} Corresponding distributions of the estimation errors
$\Delta H=H-H_{\mathrm{theor}}$ and
$\Delta\log D=\log(D/D_{\mathrm{theor}})$ for
$dt=0.02$ and $dt=17$, respectively, showing positive and negative
correlations between the errors.
\textbf{(C,G)} Joint distributions of $H$ and the predicted
effective diffusion coefficient $\log D_{\mathrm{eff}}$ for
$dt=0.02$ and $dt=17$, respectively. The units of $D_{\mathrm{eff}}$ are
($\mu\mathrm{m}^2\,\mathrm{s}^{-1}$) and are independent of $H$.
\textbf{(D,H)} Corresponding distributions of
$\Delta H=H-H_{\mathrm{theor}}$ versus
$\Delta\log D_{\mathrm{eff}}
=\log(D_{\mathrm{eff}}/D_{\mathrm{eff,theor}})$ for
$dt=0.02$ and $dt=17$, respectively.
The colorbars indicate the number of trajectories in each bin.
Cyan solid lines in (B,D,F,H) show linear fits to the corresponding
estimation errors.}
    \label{fig:degeneracy}
\end{figure}

Figure~\ref{fig:degeneracy} shows the coupling, and its removal, for short trajectories
simulated with fixed $H=0.7$ and $\log D=-1$. At $\Delta t=0.02$\,s the
estimates spread along a narrow ridge in
$(H,\log D)$ [Fig.~\ref{fig:degeneracy}(A)]: a
single true point is smeared over $0.3\lesssim H_{\mathrm{pred}}\lesssim0.9$
and more than two decades in $D$. The errors are almost perfectly correlated,
with a slope of $+4.0$\,dex per unit $H$ [Fig.~\ref{fig:degeneracy}(B)]. At
$\Delta t=17$\,s, close to the hemocyte frame interval, the ridge reverses
[Fig.~\ref{fig:degeneracy}(E)] and the slope becomes $-1.9$\,dex per unit $H$
[Fig.~\ref{fig:degeneracy}(F)], the sign we previously observed between local
estimates of $H$ and $D$ for hemocytes~\cite{korabel2022hemocytes}. Both
slopes agree with the predicted pivot $\tau^*\approx0.5\,\Delta t$
(Supplementary Sec.~S2). The correlation therefore characterises the
measurement, not the motion, and in twenty-frame windows the correlated error
in $\log D$ is large enough to mask real changes in $D$. In
$(H,\log D_{\mathrm{eff}})$ the same windows form a
compact cloud centred on the true values at both frame intervals
[Fig.~\ref{fig:degeneracy}(C,G)], and the error correlation vanishes
[Fig.~\ref{fig:degeneracy}(D,H)]. The residual slope, $\approx0.06$\,dex per
unit $H$, is less than 3\% of the original, and the errors in
$\log D_{\mathrm{eff}}$ lie within $\pm0.25$\,dex, against more than two decades
for $\log D$ at $\Delta t=0.02$\,s and more than one decade at
$\Delta t=17$\,s. The scatter that remains in $H_{\mathrm{pred}}$ is the
uncertainty of $H$ alone and no longer contaminates the amplitude.

We applied this approach to hemocytes in control, \textit{LanB1}-deficient and
collagen-IV-deficient embryos, imaged at $\Delta t=17.6$\,s per frame and
tracked in three dimensions (experimental details in
Ref.~\cite{korabel2022hemocytes}). Control and \textit{LanB1}-deficient embryos
were each recorded for $\approx5.9$\,h as six consecutive 200-frame movies,
and further embryos of all three genotypes were recorded for two movies
(Methods). To follow the motility of each cell through development, we
analysed every track in windows of 20 frames ($N=19$ steps, 334\,s), sliding
along the trajectory by one frame at a time. A window of this length is short
enough to resolve changes in behaviour along a track, but it carries limited
information, so every estimate made from it must come with its own
uncertainty. For each window, $D_{\mathrm{eff}}$ was obtained from the network
output by the fixed transformation of Eq.~\eqref{eq:Deff}; its uncertainty,
measured on simulated windows at the hemocyte frame interval, is
$0.08$--$0.09$\,dex across the range of $H$ (Methods). In what follows,
$D_{\mathrm{eff}}$ is the temperature-like measure of how strongly a cell
moves, and $H$ the measure of its directional memory.

\subsection*{PINN reproduces the hemocyte dynamics}

Before applying the PINN analysis to the hemocyte trajectories, we investigated whether the FBM model, estimated by PINN, could reproduce hemocyte dynamics. To test whether the window estimates describe the actual motion of the cells, we compared them with the frame-to-frame speed, measured directly from the positions without any model. 
\begin{figure*}[t]
  \centering
  \includegraphics[width=\textwidth]{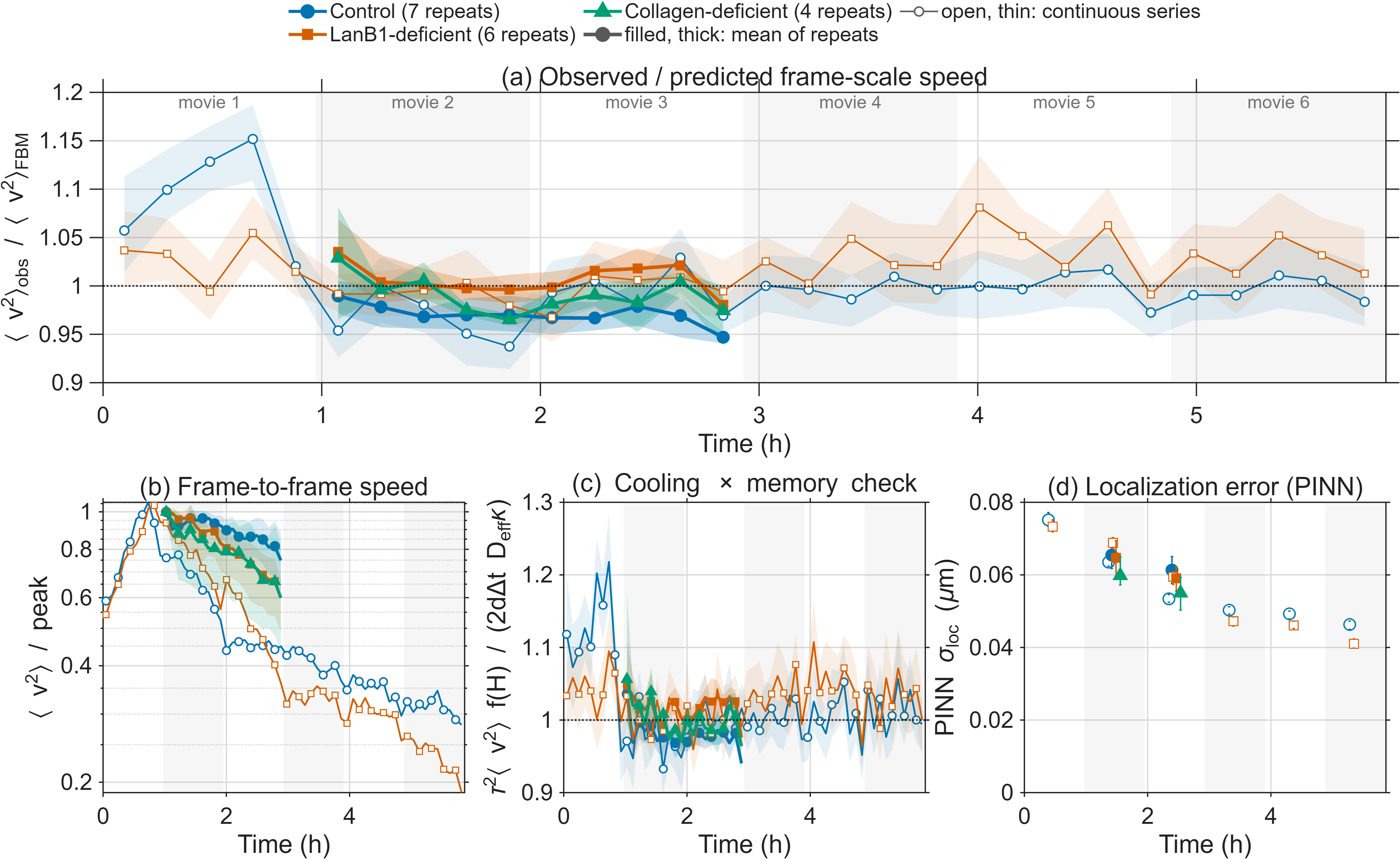}
  \caption{\textbf{Fractional Brownian motion estimated by the PINN reproduces the hemocytes dynamics.}
  Control (blue circles), \textit{LanB1}-deficient (vermillion squares) and collagen-deficient embryos
  (green triangles) hemocytes. Filled symbols and thick lines, mean of
  independent embryos (control $n=7$, \textit{LanB1}-deficient $n=6$, collagen-deficient $n=4$), 
  shown in bins covered by at least two embryos; open symbols and thin lines,
  single continuous six-movie recording. Grey bands mark alternate 200-frame
  movies.
  (a)~Ratio of observed to FBM-predicted mean squared frame-to-frame speed,
  $\langle v^{2}\rangle_{\mathrm{obs}}/\langle v^{2}\rangle_{\mathrm{FBM}}$
  [Eq.~\eqref{eq:v2fbm}], in 704\,s bins; embryo means weight each embryo
  equally. Shading, 95\% bootstrap confidence interval (over embryos for
  means, over tracks for continuous recordings); dotted line, perfect
  agreement.
  (b)~$\langle v^{2}\rangle$ normalized to its own peak (maximum of a
  three-bin moving mean), in 352\,s bins. Embryo means are estimated with one
  multiplicative offset per embryo, so that embryos entering or leaving the
  average do not shift the curve; shading, 95\% bootstrap interval over
  embryos.
  (c)~Decomposition of the speed into cooling and memory,
  $\tau^{2}\langle v^{2}\rangle f(H)/(2d\,\Delta t\,D_{\mathrm{eff}}\,\kappa)$
  [Eq.~\eqref{eq:v2deff} without the noise term], in the bins of (b); dotted
  line, exact decomposition. The ratio stays close to~1 while
  $\langle v^{2}\rangle$ falls several-fold.
  (d)~Localization error estimated by the network, per movie: median over
  tracks (95\% bootstrap interval over tracks) for continuous recordings and
  mean over embryos (95\% bootstrap interval over embryos) for embryo means.}
  \label{fig:speedchecks}
\end{figure*}
For FBM in $d=3$ dimensions with static localization error and motion blur,
the mean squared displacement over one frame, $\tau=\Delta t=17.6$\,s, is
expressed through the frame-to-frame speed
$v=|\mathbf r(t+\tau)-\mathbf r(t)|/\tau$, computed from the measured
positions of consecutive frames, with $\langle v^2\rangle$ its mean over all
pairs of consecutive frames~\cite{backlund2015chromosomal}:
\begin{equation}
\tau^{2}\langle v^{2}\rangle = 2dD\,\tau^{2H}\,\kappa(H,R) + 2d\,\sigma_{\mathrm{loc}}^{2},
\qquad
\kappa(H,R)=\frac{(1+R)^{2H+2}+(1-R)^{2H+2}-2-2R^{2H+2}}{R^{2}\,(2H+1)(2H+2)},
  \label{eq:v2fbm}
\end{equation}
where $R$ 
is the fraction of each frame interval during which
the camera is exposed. The blur factor $\kappa$ equals 1 for instantaneous
sampling and $1-R/3$ for Brownian motion ($H=0.5$). Written with the
effective diffusivity, the same expression separates the amplitude of the
motion from its memory,
\begin{equation}
  \tau^{2}\langle v^{2}\rangle =
  \frac{2d\,\tau\,D_{\mathrm{eff}}\,\kappa(H,R)}{f(H)} + 2d\,\sigma_{\mathrm{loc}}^{2},
  \qquad f(H)=\left[\det R(H)\right]^{1/N},
  \label{eq:v2deff}
\end{equation}
with $N=19$. The factor $f(H)$ is the fraction of the variance of a step that
cannot be predicted from the preceding steps. It equals 1 for Brownian motion
and is smaller for persistent motion ($f=0.96$ at $H=0.63$ and $0.76$ at
$H=0.78$), so a persistent cell covers more space per frame than a Brownian
cell at the same $D_{\mathrm{eff}}$.

At every time point of a track we combined the network estimates of all
windows containing that point ($H$ linearly, $D$ in $\log$ scale) and evaluated
Eq.~\eqref{eq:v2fbm}. The prediction was averaged over the time points of each
track in a bin and then over the tracks with matching observed speeds, and the
ratio $\langle v^{2}\rangle_{\mathrm{obs}}/\langle v^{2}\rangle_{\mathrm{FBM}}$
was formed from the two averages. The prediction is thus smoothed over
$20\Delta t\approx6$\,min, while the observed speed is a single-frame quantity.
The prediction depends on $H$ and $D$ only through $D\tau^{2H}$ at
$\tau=\Delta t$, the expected one-frame displacement. This combination is well
determined even though $H$ and $D$ individually are not, because their errors
compensate: an overestimate of $H$ comes with an underestimate of $\log D$.
Inserting the point estimates into Eq.~\eqref{eq:v2fbm} therefore biases the
prediction only slightly.

The observed speed agrees with the prediction to within a few percent in all
three genotypes [Fig.~\ref{fig:speedchecks}(a)]. 
The agreement holds while $\langle v^{2}\rangle$ falls several-fold
[Fig.~\ref{fig:speedchecks}(b)], and the decomposition of Eq.~\eqref{eq:v2deff} also holds.  That is the speed divided by
$D_{\mathrm{eff}}\kappa/f(H)$ stays close to~1 throughout the time
[Fig.~\ref{fig:speedchecks}(c)]. Because $D\tau^{2H}=D_{\mathrm{eff}}\tau/f(H)$
exactly at $\tau=\Delta t$, panels (a) and (c) express the same agreement, the
second without the localization term, which is very small here so we neglect it [Fig.~\ref{fig:speedchecks}(d)]. Mildly non-Gaussian increments and variation of speed within a window account for the
remaining scatter. Stationary FBM with the inferred $H$
and $D$ thus reproduces the short-time kinetics of hemocytes throughout
development, and the frame-scale speed is fully accounted for by
$D_{\mathrm{eff}}$ and $H$.

The largest departure occurs in the first hour of the continuous control
recording. 
The mean of seven independent control embryos shows no excess in the same bins.
We attribute the excess to a transient, embryo-specific episode of directed
migration, not to a general feature of early dispersal, and do not interpret
$D_{\mathrm{eff}}$ quantitatively in that interval.

\subsection*{Hemocytes warm, then cool; cells in \textit{LanB1}-deficient embryos cool faster and lose their memory}

\begin{figure}[t]
  \centering  \includegraphics[width=\columnwidth]{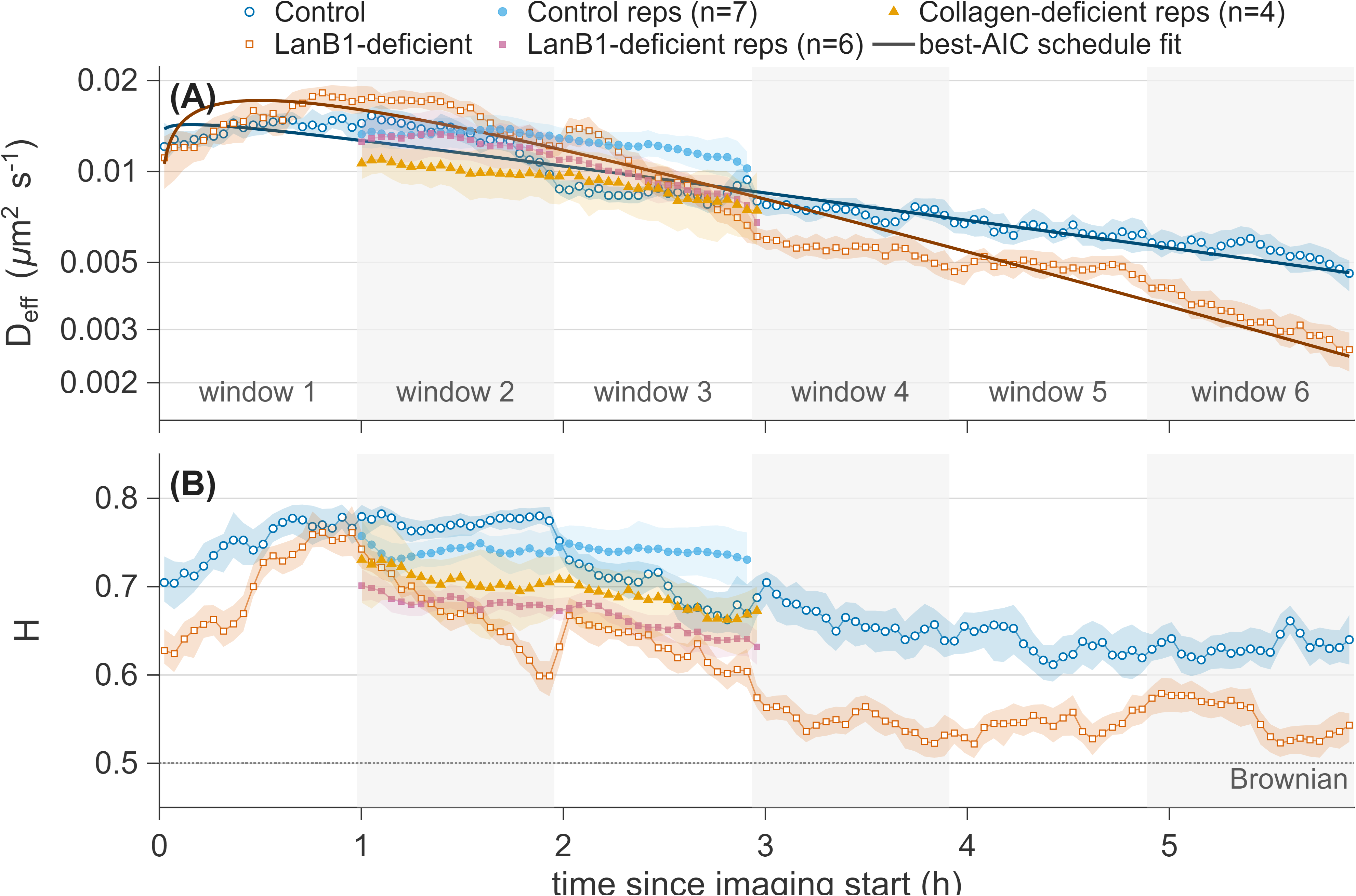}
  \caption{\textbf{Warming and cooling of hemocyte motility.}
  (A)~Effective diffusivity $D_{\mathrm{eff}}$ [Eq.~\eqref{eq:Deff}] and
  (B)~Hurst exponent $H$, in 176\,s bins (10 frames) of the time since the
  start of imaging. Symbol shape encodes genotype: control (circles), \textit{LanB1}-deficient
  (squares) and collagen-deficient embryos (triangles).
  Filled symbols, continuous six-movie recordings of control and \textit{LanB1}-deficient
  embryos; each point is the median over tracks with a 95\% bootstrap interval
  (shaded). Open symbols, mean over independent recordings: control ($n=7$)
  and \textit{LanB1}-deficient ($n=4$), each spanning movies~1-2, and collagen-deficient embryos
  ($n=4$), two recordings spanning movies~1-2 and two spanning movies~2-3. Each open point is the mean of the per-recording medians, with a 95\%
  interval from a two-level bootstrap over recordings and over tracks within
  each recording. A bin is shown when at least 2 recordings contribute, each
  with at least 5 tracks.
  Dashed lines, best-fitting cooling schedules for the continuous recordings,
  selected by AIC among exponential, power-law, logarithmic and rise--decay
  forms. Grey and white bands mark consecutive 200-frame movies (58.7\,min
  each). Dotted line in (B), Brownian motion, $H=0.5$. Estimates come from
  20-position windows ($\Delta t=17.6$\,s). Symbols are drawn every second
  bin.}
  \label{fig:schedule}
\end{figure}

To follow the population through development, we averaged $\log D_{\mathrm{eff}}$
and $H$ over the windows of each track in 176\,s bins and took the median
across tracks, with a bootstrap confidence interval
[Fig.~\ref{fig:schedule}(A,B)].
The time course has two phases. During the first hour both genotypes warm and
gain persistence, as the hemocytes disperse through the embryo.
$D_{\mathrm{eff}}$  and $H$ rises in all embryos. Thereafter different genotypes cool at different rates. Control hemocytes cool slowly. $D_{\mathrm{eff}}$ falls
about 3-fold from its maximum , to $\approx0.005\,\mu\mathrm{m}^2\,\mathrm{s}^{-1}$
at 5.9\,h. Hemocytes in \textit{LanB1}-deficient embryos cool about 6-fold, to
$\approx0.003\,\mu\mathrm{m}^2\,\mathrm{s}^{-1}$. Cells in \textit{LanB1}-deficient embryos are therefore
hotter than control cells for the first $\approx2.7$\,h, by up to $0.2$\,dex,
and colder afterwards.

\begin{figure*}[t]
  \centering   
  \includegraphics[width=\columnwidth]{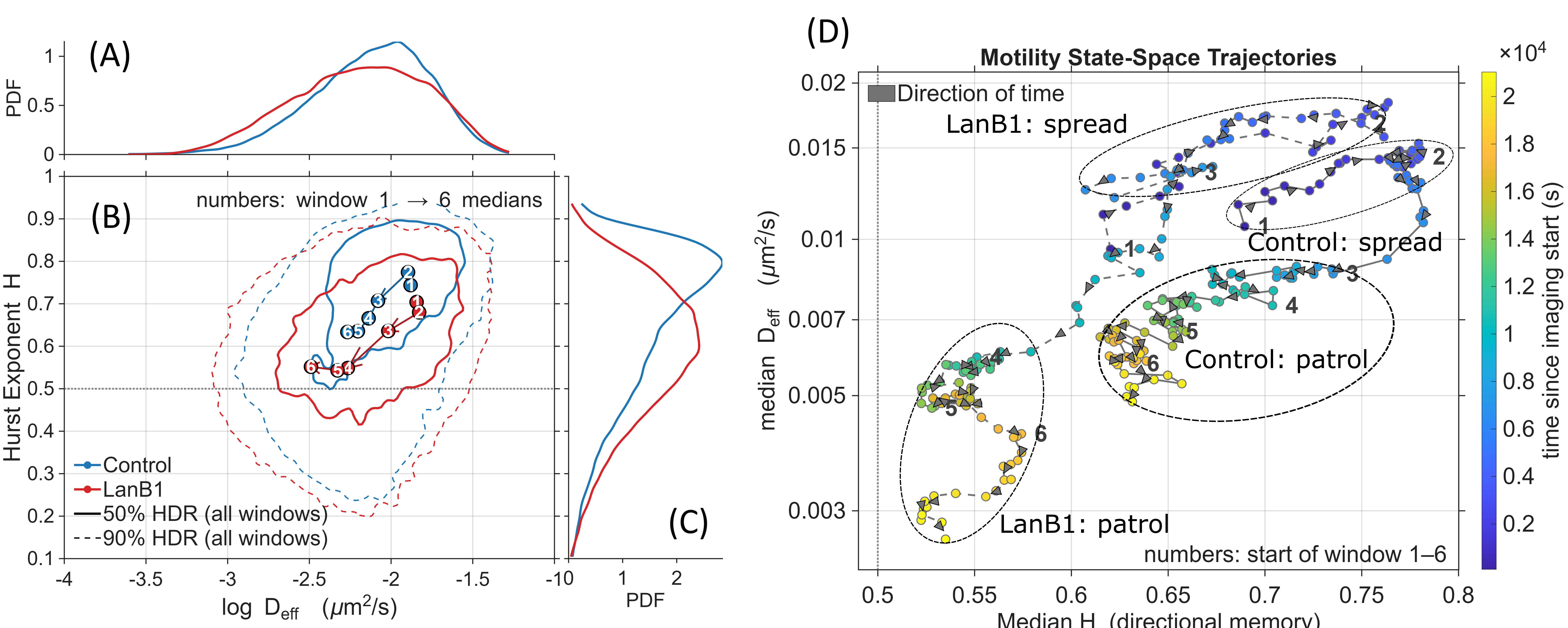}
  \caption{\textbf{Trajectories through memory and temperature.}
  \textbf{(A,C)}~Marginal distributions of $\log D_{\mathrm{eff}}$ and $H$, pooled over
  windows and tracks. \textbf{(B)}~Joint distribution of $(\log D_{\mathrm{eff}},H)$:
  solid and dashed contours enclose the 50\% and 90\% highest-density regions;
  numbered circles, medians of movies 1--6, joined in time order.
  \textbf{(D)}~Population medians in 176\,s bins, coloured by time since the start of
  imaging; solid, control; dashed, \textit{LanB1}-deficient embryos; arrowheads give the direction of
  time; numbers mark the first bin of each movie. Dashed outlines enclose the
  bins of the spread (movies 1--2) and patrol (movies 4--6) phases. Dotted
  lines, Brownian motion ($H=0.5$).}
  \label{fig:statespace}
\end{figure*}

Persistence follows a different course in the two genotypes
[Fig.~\ref{fig:schedule}(B)]. Control hemocytes hold $H\approx0.78$ until
$\approx1.9$\,h, then lose persistence gradually and settle at $H\approx0.63$
from $\approx4$\,h, well above the Brownian value. Hemocytes in \textit{LanB1}-deficient embryos begin to
lose persistence as soon as they reach their maximum, at $\approx1$\,h; they
fall to $H\approx0.61$ by 1.9\,h and remain at $H\approx0.52$--$0.57$ after
3\,h, close to Brownian motion. Cells in \textit{LanB1}-deficient embryos are less persistent than control cells. In the plane of temperature and memory
the two genotypes therefore trace different paths: control cells cool at
nearly constant memory and lose it only late, whereas cells in \textit{LanB1}-deficient embryos lose memory and cool together [Fig.~\ref{fig:statespace}].

Equation~\eqref{eq:v2deff} separates the slowdown into two parts: cooling
(the fall in $D_{\mathrm{eff}}$) and loss of memory (the fall in $H$), which
enters through $f(H)$. 
The loss
of persistence alone, from $H\approx0.78$ to $0.63$ in control and from $0.76$
to $0.55$ in \textit{LanB1}-deficient embryos, lowers the frame-scale speed by $\approx1.25$-fold in
both genotypes. The model-free $\langle v^{2}\rangle$ falls $\approx3.7$-fold
in control and $\approx5.3$-fold in \textit{LanB1}-deficient embryos over the same period
[Fig.~\ref{fig:speedchecks}(b)], so most of the slowdown is cooling. Speed alone would mix the two contributions; $D_{\mathrm{eff}}$ isolates the
amplitude of the motion and $H$ its memory.

\begin{figure}[t]
  \centering
  \includegraphics[width=\linewidth]{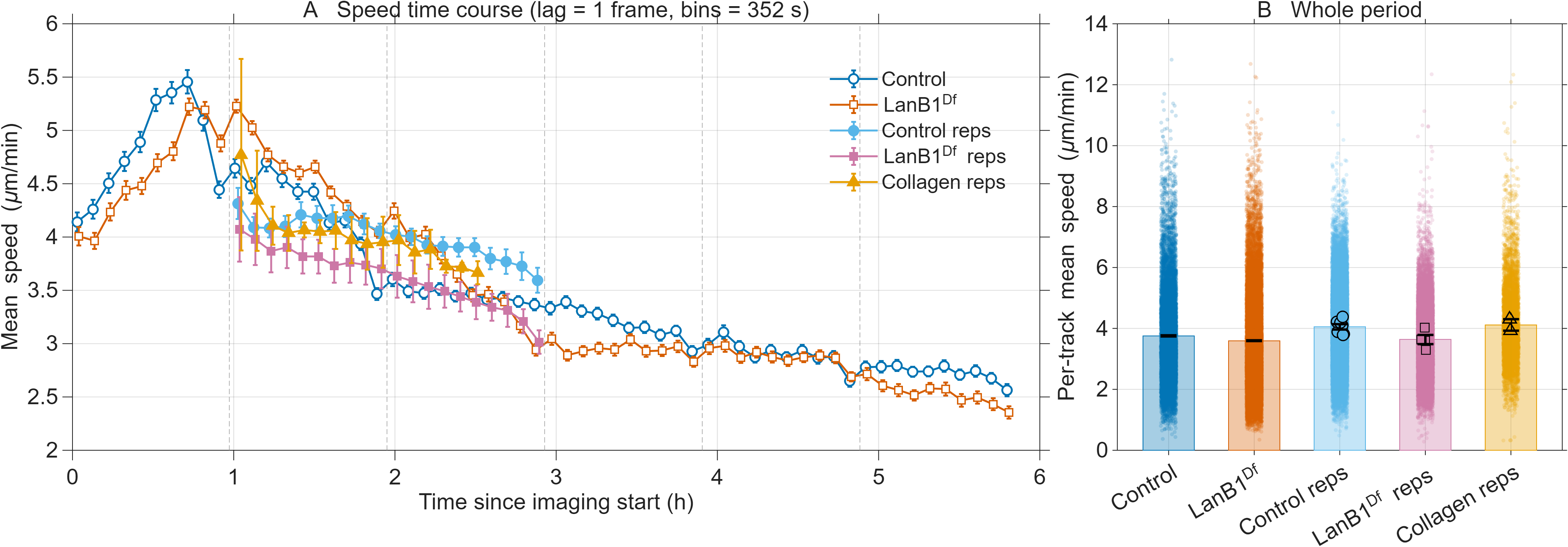}
\caption{\textbf{Hemocyte speed across development and genotypes.}
  Speed is $v=|\mathbf{r}(t+\Delta t)-\mathbf{r}(t)|/\Delta t$, computed from
  3D positions in consecutive frames ($\Delta t=17.6$\,s); pairs across
  tracking gaps are excluded. Symbol shape encodes genotype: control
  (circles), \textit{LanB1}-deficient (squares) and the collagen-deficient embryos (triangles).
  Open symbols are the continuous six-movie recordings of control and \textit{LanB1}-deficient embryos; filled symbols are means over independent recordings:
  control ($n=7$) and \textit{LanB1}-deficient embryos ($n=4$), each spanning movies~1--2, and the
  collagen-deficient embryos ($n=4$), two recordings spanning movies~1--2 and two
  spanning movies~2--3.
  (A)~Mean speed in 352\,s bins (20 frames), calculated by first averaging
  each track within a bin. For the continuous recordings, points are means
  over tracks $\pm$ SEM across tracks (at least 5 tracks per bin). For the
  independent recordings, each recording is first averaged over its tracks
  (at least 5 per bin), and points are means over recordings $\pm$ SEM
  across recordings (at least 2 recordings per bin). Dashed vertical lines
  mark the boundaries between consecutive 200-frame movies. Points are offset
  slightly in time for visibility.
  (B)~Mean speed over each dataset's full recording period. Small dots are
  per-track mean speeds (tracks with at least 5 speed samples). Large symbols
  are per-recording means. Bars and error bars are the mean $\pm$ SEM, across
  tracks for the continuous recordings and across recordings for the
  independent recordings. The datasets cover different periods (movies~1--6,
  1--2 or 1--3), so the bars are not time-matched.}
  \label{fig:speed}
\end{figure}

The cooling is visible without the network. Frame-to-frame speeds measured
directly from the positions, averaged first within each track and then across
tracks, show the same two phases [Fig.~\ref{fig:speed}(A)]. During the first
$\approx0.7$\,h control cells accelerate from $\approx4.1$ to
$\approx5.4\,\mu\mathrm{m\,min^{-1}}$ and cells in \textit{LanB1}-deficient embryos from $\approx4.0$ to
$\approx5.2\,\mu\mathrm{m\,min^{-1}}$. Both then slow down steadily, to
$\approx2.6$ (control) and $\approx2.35\,\mu\mathrm{m\,min^{-1}}$ (\textit{LanB1}-deficient embryos) by
the end of imaging, roughly a twofold fall from the peak, or a 4--5-fold fall
in $\langle v^{2}\rangle$. Averaged over the whole recording, the per-track mean
speed differs little between the genotypes ($3.75\pm0.015$ against
$3.60\pm0.012\,\mu\mathrm{m\,min^{-1}}$, mean $\pm$ s.e.m.; see Fig.~\ref{fig:speed}(B)).
Several further checks show that the trends are not the artifacts of the
measurements. The confidence intervals of the medians are much narrower than
the changes they track. Localization noise, which is negligible here, could
only add a constant floor to $\langle v^{2}\rangle$
[Eq.~\eqref{eq:v2deff}] and slow the apparent decline, never create it. Different 
cooling scenarios are barely distinguishable in the raw diffusion coefficient, because in
$\mu\mathrm{m}^2\,\mathrm{s}^{-2H}$ a falling $H$ masks a falling amplitude (see Supplementary Fig. S6),
but the apparent diffusivity at a fixed lag, $D\tau^{2H-1}$, falls differently in both
genotypes. 

The time course matches the developmental programme of embryonic hemocytes: a
hot, persistent dispersal phase, in which the cells spread through the embryo,
followed by gradual cooling into a slower, less persistent patrolling phase
suited to local surveillance and matrix deposition. In the movies of every genotype, the initial decline is indistinguishable between
control, \textit{LanB1}-deficient and collagen-deficient embryos. Later, the continuous \textit{LanB1}-deficient embryos recording cools further and approaches Brownian
motion more closely than the control, which points to a role for laminin in
sustaining an active, persistent search. These late differences rest on one
continuous recording per genotype and are tested at the level of single cells
in the next section.
For a compact description we fitted four schedules to $\ln D_{\mathrm{eff}}(t)$:
exponential, power-law, logarithmic ($D_{\mathrm{eff}}\propto1/\ln t$, the
schedule under which simulated annealing is guaranteed to
converge~\cite{geman1984stochastic}), and rise--decay,
$D_{\mathrm{eff}}\propto t^{\beta}e^{-t/\tau}$. Only the rise--decay form
captures the initial warming, and the Akaike information criterion prefers it
in both genotypes. It gives a maximum at
$t_{\mathrm{peak}}=0.24$\,h and a cooling time $\tau=4.5$\,h in control
embryos, against $0.52$\,h and $2.13$\,h in \textit{LanB1}-deficient embryos. 
Because the fits do not model the steps at movie boundaries, we use them as
descriptive summaries, not as tests between mechanisms.

\subsection*{Fast cells remember: speed--persistence coupling is lost in \textit{LanB1}-deficient embryos}

\begin{figure*}[t]
  \centering
  \includegraphics[width=\textwidth]{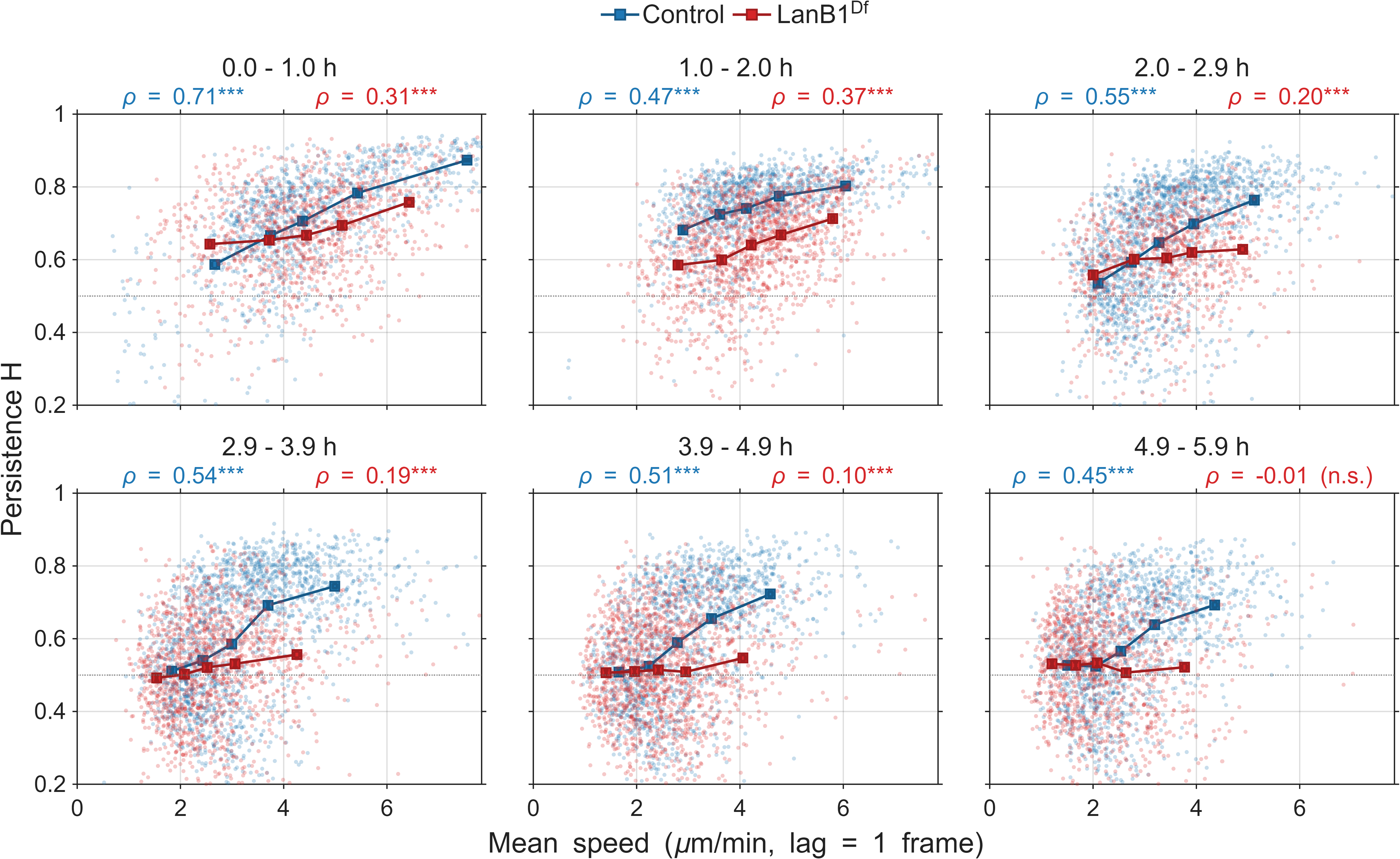}
  \caption{\textbf{Speed--persistence coupling is preserved in control
  hemocytes and lost in \textit{LanB1}-deficient embryos.} Each panel corresponds to one movie
  ($\approx1$\,h). Dots: individual track segments, plotting the mean
  frame-to-frame speed against the mean Hurst exponent $H$ of windows that
  passed the goodness-of-fit test ($p>0.05$); segments required at least
  five speed samples and five $H$ estimates. Squares: means of speed
  quintiles, with error bars giving the s.e.m.\ in both coordinates. Dotted
  line: Brownian motion ($H=0.5$). $\rho$: Spearman correlation per genotype
  (***\,$p<0.001$; n.s., not significant). Numbers of segments per panel,
  control/\textit{LanB1}-deficient: 1174/1089, 1126/1450, 1711/924, 1692/1414, 1484/1601,
  1391/1256.}
  \label{fig:Hspeed}
\end{figure*}

Speed and persistence are coupled in many migrating cells: faster cells turn
less, through the advection of polarity cues by actin flow~\cite{maiuri2015actin}.
We asked whether this coupling holds for hemocytes and whether it survives
cooling. For each track segment and each movie, we paired the mean speed with
the mean $H$ of its windows, keeping only windows that passed the
goodness-of-fit test ($p>0.05$) [Fig.~\ref{fig:Hspeed}].

In control embryos the coupling is strong at every stage. The Spearman
correlation between speed and $H$ is $\rho=0.71$ in the first hour and stays
between $0.45$ and $0.55$ from the second hour onwards ($p<10^{-10}$
throughout). In the first hour, $H$ rises from $\approx0.6$ for the slowest
fifth of the cells to $\approx0.87$ for the fastest; in the last hour the rise
is smaller but still clear, from $\approx0.53$ to $\approx0.69$. Cooling
therefore shifts the control population to lower speeds along a preserved
speed--persistence relation: slow cells are nearly Brownian, but fast cells
still keep their direction.

In \textit{LanB1}-deficient embryos the coupling is weaker from the start ($\rho=0.31$ and
$0.37$ in the first two hours). It then decays steadily, to $\rho=0.20$, $0.19$
and $0.10$ in the following hours, and vanishes in the last hour
($\rho=-0.01$, $p=0.6$), where $H$ stays at $\approx0.51$--$0.53$ across the
whole speed range. Late in development the two genotypes coincide for the
slowest cells, where both are close to $H=0.5$, and diverge for the fast cells:
hemocytes in \textit{LanB1}-deficient embryos that move as fast as persistent control cells do so without
directional memory. The late loss of persistence in hemocytes embryos is thus a loss of
memory in the fastest cells, not a shift of the population towards slow,
Brownian motion. Laminin is a component of the basement membranes along which
embryonic hemocytes migrate, and hemocytes themselves deposit
it~\cite{sanchez2017drosophila}. The data are consistent with the basement
membrane providing the guidance that converts speed into persistence.

\subsection*{An annealed search}

These observations can be interpreted as an \emph{annealed search}. If each
hemocyte moves with an effective temperature
$T_{\mathrm{eff}}=\gamma D_{\mathrm{eff}}/k_B$ in a landscape of attractive
sites, an early high temperature lets the population cross weak barriers and
spread through the embryo. As the temperature falls, random motion no longer
overwhelms weak attractions, which increasingly determine where the cells
settle and patrol. At constant friction the relative cooling
$T_{\mathrm{eff}}(t)/T_{\mathrm{eff}}(0)$ equals
$D_{\mathrm{eff}}(t)/D_{\mathrm{eff}}(0)$ and does not depend on the unknown
$\gamma$. In this picture, control hemocytes follow a slow schedule and keep
most of their directional memory while patrolling, whereas \textit{LanB1}-deficient hemocytes
are heated more strongly, cooled faster and further, and lose most of their
memory. The friction, however, is not guaranteed to be constant. Deposition of
basement membrane and compaction of tissue could raise $\gamma$ and lower
$D_{\mathrm{eff}}$ without any change in the cells' activity, and in a laminin
mutant the matrix itself is altered. The two explanations make different
predictions, which we set out in the Discussion.


\section*{Methods}

\subsection*{Data}
Embryo preparation, selective plane illumination imaging and three-dimensional
tracking of hemocytes were as described in Ref.~\cite{korabel2022hemocytes}.
Positions were recorded at $\Delta t=17.6$\,s per frame with an exposure
fraction $R\approx0.05$. 
Control and \textit{LanB1}-deficient embryos were each recorded continuously for six consecutive
200-frame movies ($\approx5.9$\,h). In addition, independent recordings of two
consecutive movies were analysed from seven control embryos (movies 1--2), six
\textit{LanB1}-deficient embryos (four spanning movies 1--2 and two spanning movies 2--3) and
four collagen-deficient embryos (two spanning movies 1--2 and
two spanning movies 2--3). 
Tracks were split at missing frames, and contiguous segments of at least 20
positions entered the window analysis. \textit{LanB1}-deficient embryos are homozygous for a deletion of the \textit{LanB1} locus \cite{urbano2009drosophila}  while Collagen-IV deficient embryos are homozygous for the small chromosomal deletion Df(2L)BSC172 which removes the genes encoding the Collagen-IV components Col4a1 and Viking \cite{matsubayashi2017moving}.

\subsection*{Observation model}
For a window of $N+1$ positions we write the displacements on axis $c$ as
$\delta^{(c)}_i=y^{(c)}_{i+1}-y^{(c)}_i$, $i=1,\dots,N$, with $N=19$. For
fractional Brownian motion with
$\langle\Delta x^2(t)\rangle=2Dt^{2H}$ per axis, the displacements are
fractional Gaussian noise with correlation function
\begin{equation}
  \varrho(k)=D\,\Delta t^{2H}\bigl(|k+1|^{2H}-2|k|^{2H}+|k-1|^{2H}\bigr).
  \label{eq:fgn}
\end{equation}
The measured positions carry a static localization error,
$\mathbf y_k=\mathbf X(t_k)+\boldsymbol\varepsilon_k$, with
$\boldsymbol\varepsilon_k\sim\mathcal N(\mathbf 0,\sigma_{\mathrm{loc}}^2\mathbb I_3)$
independent between frames and axes, and each position is averaged over the
exposure $t_E=R\Delta t$ (box illumination). The covariance of the observed
displacements is then
\begin{equation}
  \Sigma_{ij}=\varrho_R(|i-j|)
   +\sigma_{\mathrm{loc}}^2\bigl(2\delta^{\mathrm K}_{ij}-\delta^{\mathrm K}_{|i-j|,1}\bigr),
  \label{eq:Sigma}
\end{equation}
where $\varrho_R$ is the correlation function of the blurred process, obtained
from its mean squared displacement~\cite{backlund2015chromosomal}
$M(n)=2D\Delta t^{2H}\bigl[(n+R)^{p}+(n-R)^{p}-2n^{p}-2R^{p}\bigr]/[R^2(p-1)p]$,
$p=2H+2$, as $\varrho_R(k)=\tfrac12[M(k+1)-2M(k)+M(|k-1|)]$. At lag one this
gives the blur factor $\kappa(H,R)=M(1)/(2D\Delta t^{2H})$ of
Eq.~\eqref{eq:v2fbm}. The three axes are independent and share $\Sigma$. We
define the signal-to-noise ratio as
$\mathrm{SNR}=\sqrt{2D\Delta t^{2H}}/\sigma_{\mathrm{loc}}$. The background to
Eq.~\eqref{eq:Sigma} is given in Supplementary Sec.~S1.

\subsection*{Network}
\paragraph{Inputs.} The displacements of a window are normalized by their own
root mean square,
$s^2=(3N)^{-1}\sum_{c}\sum_{i}\bigl(\delta^{(c)}_i\bigr)^2$, and the $3N$
values $\delta^{(c)}_i/s$ are the input. The network is given neither
$\Delta t$ nor the units of the coordinates, so one trained network applies at
any frame rate and over any range of $D$. Two scalars are appended: a flag
stating whether $\sigma_{\mathrm{loc}}$ is known, and
$\log_{10}(\sigma_{\mathrm{loc}}/s)$ when it is (a fixed value otherwise).

\paragraph{Architecture.} A shared trunk of four fully connected layers (256,
256, 128 and 64 units, SiLU activations) feeds three two-layer heads. The first
head returns $H$, through a rescaled sigmoid restricted to $0.02<H<0.98$, and
its standard deviation $\sigma_H$. The second head returns a bounded correction
$r$ ($|r|\le0.87$\,dex) and the standard deviation $\sigma_{\log D}$. The scale
removed by the normalization is restored analytically,
\begin{equation}
  \log_{10}D=r+\log_{10}s^2-2H\log_{10}\Delta t-\log_{10}2-\log_{10}\kappa(H,R),
  \label{eq:logD}
\end{equation}
so the network learns a correction of order one decade rather than the
magnitude of $D$. The third head returns $\sigma_{\mathrm{loc}}/s$ (the localization error due to the finite resolution of the microscope), bounded
between $10^{-2}$ and $1$; when $\sigma_{\mathrm{loc}}$ is supplied it replaces
this output.

\paragraph{Loss.} The loss is
\begin{equation}
  L=w(e)\,L_{\mathrm{phys}}+L_H+L_{\log D}+\lambda_\sigma L_\sigma .
  \label{eq:total}
\end{equation}
The physics term is the exact Gaussian negative log-likelihood of the observed
displacements,
\begin{equation}
  L_{\mathrm{phys}}=\frac{1}{3N}\Bigl[\frac12\sum_{c=1}^{3}
   \boldsymbol\delta^{(c)\mathsf T}\Sigma^{-1}\boldsymbol\delta^{(c)}
   +\frac32\ln\det\Sigma+\frac{3N}{2}\ln2\pi\Bigr],
  \label{eq:phys}
\end{equation}
with $\Sigma$ from Eq.~\eqref{eq:Sigma} built from the predicted parameters and
evaluated by Cholesky factorization. It uses no ground truth. $L_H$ and
$L_{\log D}$ are heteroscedastic Gaussian losses,
$\langle(\hat\theta-\theta)^2/2\sigma_\theta^2+\ln\sigma_\theta\rangle$, whose
minimum with respect to $\sigma_\theta$ lies at the realized error of each
window, so the uncertainties are learned with the point estimates and are not
rescaled afterwards. $L_\sigma$ is a squared error on
$\log_{10}(\sigma_{\mathrm{loc}}/s)$, evaluated only where $\sigma_{\mathrm{loc}}$
was not supplied, with $\lambda_\sigma=(\ln10)^2$. The weight
$w(e)=\min(1,e/500)$ ramps in the likelihood term over the first 500 epochs
(Supplementary Sec.~S1).

\paragraph{Training.} A new batch of 512 three-dimensional trajectories of 20
positions was simulated at every epoch. For each trajectory, $H$ was drawn
uniformly in the range $(0.04,0.96)$, $\log D$ uniformly in $(-5,1)$, $\Delta t$
uniformly in $(0.01,0.03)$\,s and the static localization error was set by the signal-to-noise ratio,
$\sigma_{\mathrm{loc}}=s/\mathrm{SNR}$, with SNR drawn log-uniformly in
$(2,50)$, where $s=\sqrt{2D\Delta t^{2H}}$ is the standard deviation of a single
step along one axis. The
displacements were generated from the exact covariance by Cholesky
factorization, with the exposure fraction drawn within $\pm10\%$ of $R=0.05$,
and Gaussian localization noise was added to the positions. The value of
$\sigma_{\mathrm{loc}}$ was supplied to the network for a random half of each
batch. We used AdamW (learning rate $3\times10^{-4}$, weight decay $10^{-4}$)
with gradient clipping for 12\,000 epochs, evaluated the network every 100
epochs on eight independent validation batches, halved the learning rate when
the validation loss reached a plateau, and kept the network with the lowest
validation loss. Performance, calibration and the dependence on window length and SNR are
reported in Supplementary Sec.~S1.

\subsection*{Inference along trajectories}
Each contiguous segment was analysed in windows of 20 positions sliding by one
frame, with the exposure fraction checked against the value used in training.
The estimates of all windows covering a time point were combined: $H$ and
$\log_{10}D$ as means weighted by $1/\sigma^2$, and their uncertainties as the
correspondingly weighted mean of the window uncertainties, because overlapping
windows share most of their displacements and their errors are almost perfectly
correlated. 

\subsection*{Effective diffusivity calculation and its uncertainty}
To calculate the effective diffusion coefficient Eq.   \ref{eq:Deff}
we need to calculate the $N\times N$ correlation matrix $R(H)$ of fractional Gaussian noise. $\log \det R(H)$ was tabulated by Cholesky factorization on a grid
$H\in[0.02,0.98]$ and interpolated. $D_{\mathrm{eff}}$ has units of
$\mu\mathrm m^2\,\mathrm s^{-1}$ at every $H$, equals $D$ at $H=1/2$, and
$2\Delta t\,D_{\mathrm{eff}}$ is the geometric mean of the variances of the
parts of successive steps that cannot be predicted from the preceding steps.

Because the network does not return the covariance between the errors of $H$
and $\log D$, the uncertainty of $\log D_{\mathrm{eff}}$ was measured
directly. We simulated 4\,000 three-dimensional FBM
tracks of 40 positions at $\Delta t=17.6$\,s, with $H$ uniform in
$(0.45,0.90)$, per-axis step size uniform in $(0.4,1.2)$\,\textmu m and SNR=20 (we also tested SNR
log-uniform in $(5,50)$), analysed them with the same network and inference script, and tabulated the standard deviation of the error of
$\log D_{\mathrm{eff}}$ in bins of the predicted $H$. The error is
$0.080$--$0.092$\,dex over the whole range, while the errors of $H$ and
$\log D$ are correlated with coefficient $-0.8$ to $-0.9$.  

\subsection*{Time courses}
Each window was assigned to a 176\,s bin (10 frames) by its time stamp, with
movies placed on a common clock from their frame offsets. Within each track and
bin, the estimates of $H$ and $\log D_{\mathrm{eff}}$ were averaged, giving
one value per track and bin. For a single recording we report the median of the per-track values with a 95\%
percentile interval from 500 bootstrap resamples of tracks, for bins with at
least five tracks. For independent recordings of the same genotype we report
the mean of the per-recording medians, in bins covered by at least two
recordings with at least five tracks each, with a 95\% interval from a
two-level bootstrap over recordings and over tracks within each recording.
Treating each track as one unit gives tracks equal weight and keeps the
resampled units independent. The rise--decay, exponential, power-law and
logarithmic schedules were fitted to $\ln D_{\mathrm{eff}}(t)$ of the
continuous recordings by least squares and compared using the Akaike information
criterion.

\subsection*{Frame-scale speed}
Squared frame-to-frame speeds,
$v^2=|\mathbf r(t+\Delta t)-\mathbf r(t)|^2/\Delta t^2$, were computed from
positions in consecutive frames. At every time point the FBM prediction
$\langle v^2\rangle_{\mathrm{FBM}}$ was evaluated from Eq.~\eqref{eq:v2fbm} with
the combined network estimates. Windows with goodness-of-fit $p\le0.05$ or with
$H$ outside $[0.02,0.98]$ were excluded. Within each time bin, observed and
predicted values were averaged first per track (at least three samples of
each) and then over tracks (at least ten tracks per bin), and the ratio in
Fig.~\ref{fig:speedchecks}(a) is the ratio of these track averages; bins were
704\,s for the ratio and 352\,s for the other panels. The decomposition check of
Fig.~\ref{fig:speedchecks}(c) is the same ratio written as
$\tau^2\langle v^2\rangle f(H)/(2d\,\Delta t\,D_{\mathrm{eff}}\kappa)$
[Eq.~\eqref{eq:v2deff}]. Embryo means of the two ratios are equal-weight
averages over embryos, shown where at least two embryos contribute. Because
embryos cover different time spans, a plain average of absolute values such as
$\langle v^2\rangle$ would change whenever an embryo enters or leaves it. The
embryo mean of $\langle v^2\rangle$ was therefore obtained by fitting
$\log y_{br}=a_r+f_b$ (one offset per embryo $r$, one value per bin $b$,
$\sum_r a_r=0$) by alternating least squares; $\exp(f_b)$ is the time course of
a typical embryo, shown normalized to the maximum of its three-bin moving
average. Confidence intervals are 95\% percentile intervals from 200 bootstrap
resamples, of tracks for single recordings and of embryos for means. Mean
speeds (Fig.~\ref{fig:speed}) were computed from
$v=|\mathbf r(t+\Delta t)-\mathbf r(t)|/\Delta t$, averaging each track within a
bin first; error bars are standard errors across tracks for single recordings
and across recordings for independent recordings.



\subsection*{Data and code availability}
The training and inference scripts for the network, as well as the analysis scripts, will be available on GitHub upon publication. 


\section*{Discussion}

We have followed the motility of \textit{Drosophila} hemocytes through six
hours of embryonic development in two coordinates: an effective diffusivity
$D_{\mathrm{eff}}$, which measures how strongly a cell moves, and the Hurst
exponent $H$, which measures how much it remembers its previous direction. In
both genotypes followed for six hours (the control and \textit{LanB1}) the
time course has two phases. During the first hour the cells warm and become
more persistent. Thereafter they cool. Control hemocytes cool about threefold
and settle at $H\approx0.63$, remaining clearly persistent while they patrol.
Hemocytes in \textit{LanB1}-deficient embryos are hotter at first but cool
about sixfold, and they approach Brownian motion. Hemocytes in
collagen-IV-deficient embryos, recorded over the second and third hours, cool
over the same period but keep most of their persistence. The spread between
cells grows during development and, except during the second movie, exceeds
the error of the individual estimates, so the late embryo contains a
genuinely heterogeneous population.

These conclusions depend on two methodological steps. The first is a
calibrated uncertainty for every local estimate. In twenty-frame windows the
estimation error is comparable to the biological variation, and without it
neither the changes along a track nor the spread between cells could be
interpreted. The second is the choice of coordinate for the amplitude of
motion. The generalized diffusion coefficient $D$ carries units of
$\mathrm{L}^2\,\mathrm{T}^{-2H}$, which differ whenever $H$ differs, and in
short windows its error is almost entirely inherited from the error in $H$.
The resulting correlation between $H$ and $\log D$ is a property of the
measurement: its sign is set by the frame rate, and it is positive at
millisecond frame intervals and negative at the 17.6\,s interval used here.
Correlations of this kind between estimated exponents and diffusion
coefficients are widely reported and can even reverse the sign of the true
correlation~\cite{lanoiselee2025super}. The negative correlation we previously
observed for hemocytes~\cite{korabel2022hemocytes} belongs to the same class.
$D_{\mathrm{eff}}$ removes the problem at the level of the likelihood. Its
units are $\mu\mathrm{m}^2\,\mathrm{s}^{-1}$ for every $H$, it equals $D$ for
Brownian motion, and its error is asymptotically uncorrelated with that of
$H$. Physically, it measures the unpredictable part of each step, that is,
the part of the motion not explained by the cell's recent history. With this
transformation, the cooling of hemocytes, which is nearly invisible in $D$
because a falling $H$ masks a falling amplitude, becomes the dominant feature
of the data.

The two phases coincide with the two stages of hemocyte behaviour known from
developmental studies: dispersal from the head mesoderm along stereotyped
routes, followed by residence in the tissues, where the cells clear apoptotic
corpses and respond to wounds~\cite{wood2007drosophila,evans2014drosophila}.
Our measurements add a physical description of this transition. Dispersal is
hot and persistent, and patrolling is cooler. Control cells cool while
retaining much of their directional memory, so they continue to sweep their
surroundings in persistent runs rather than diffusing in place. This is the
pattern expected of an annealed
search~\cite{kirkpatrick1983optimization,geman1984stochastic}. A high early
effective temperature allows the population to cross weak barriers and spread
through the embryo. As the temperature then falls, weaker barriers become
strong enough to hold a cell, so that the cells settle at favourable sites and
patrol around them. The strategy differs from other standard paradigms such as
run-and-tumble motion and from L\'evy walks~\cite{harris2012generalized}, in
which the statistics of motion are stationary. It shares features with
intermittent search~\cite{benichou2011intermittent}, the alternation of
relocation and local exploration, but here the alternation unfolds over
developmental time rather than along each trajectory. We note that simulated
annealing is usually formulated for a single searcher seeking one global
minimum. Hemocytes instead form a population that must cover the whole
embryo, so the relevant outcome is the redistribution of the population, not
the convergence of individual cells.

Loss of laminin B1 alters both the effective temperature and the directional
memory of hemocytes. Hemocytes in \textit{LanB1}-deficient embryos are hotter
during dispersal, cool faster and further, and lose most of the persistence
that control cells retain. Hemocytes deposit components of the basement
membrane~\cite{matsubayashi2017moving}, and the laminin they secrete
strengthens their own migration~\cite{sanchez2017drosophila}. One
interpretation of our data is therefore that laminin is needed to sustain
directed motility during patrol, not to set the effective temperature. The
earlier observation that the \textit{LanB1} deficiency reduces the persistence
of hemocyte motion~\cite{korabel2022hemocytes} is consistent with this
interpretation, and our analysis locates the loss of persistence in the patrol
phase. A second interpretation, which we cannot fully exclude, is mechanical:
an altered matrix could change the confinement and friction experienced by the
cells, as discussed below.

Collagen~IV, the other main structural component of the basement membrane and
one that hemocytes largely deposit themselves~\cite{matsubayashi2017moving},
provides a contrast. In collagen-IV-deficient embryos the frame-scale speed is
again reproduced by fractional Brownian motion
[Fig.~\ref{fig:speedchecks}(a)], and the hemocytes cool over the period we can
follow, slightly below control cells in $D_{\mathrm{eff}}$. Their persistence,
however, stays close to that of control cells, at $H\approx0.66$--$0.72$, and
above that of \textit{LanB1}-deficient cells at the same stage
[Fig.~\ref{fig:schedule}]. 
Reducing collagen~IV therefore does not reproduce the loss of memory caused by
the loss of laminin. This points to laminin itself, rather than the integrity
of the basement membrane as a whole, as the component that converts speed into
persistence, consistent with the role of hemocyte-derived laminin in
stabilising lamellipodia and sustaining persistent
migration~\cite{sanchez2017drosophila}. Two limits apply. The collagen
recordings end before the patrol phase, where the memory of
\textit{LanB1}-deficient cells is lost most clearly, and four embryos give
little power to detect a modest change in $H$. Following collagen-deficient
embryos into the late embryo, and testing the speed--persistence coupling of
Fig.~\ref{fig:Hspeed} in them, would show whether collagen~IV is dispensable
for persistent patrol or only for its onset.

A falling $D_{\mathrm{eff}}$ is a direct observation. Interpreting it as a
falling effective temperature, $T_{\mathrm{eff}}=\gamma D_{\mathrm{eff}}/k_B$,
requires the friction $\gamma$ to remain constant. The fluids of the fly are
simple: larval hemolymph has a viscosity close to that of water,
$1.34\pm0.07$\,cP~\cite{zabihihesari2023pdms}, and the embryonic cytoplasm and
yolk respond predominantly viscously, at $\sim10^3$\,cP, on the scale of
micrometre probes and minutes~\cite{doubrovinski2017measurement}. Hemocytes,
however, do not move as particles suspended in these fluids. A cell of radius
5\,\textmu m diffusing thermally in hemolymph would have
$D\approx0.03\,\mu\mathrm{m}^2\,\mathrm{s}^{-1}$, more than the values of
$D_{\mathrm{eff}}$ we measure. The friction that limits hemocyte motion is
therefore set by adhesion and confinement within the tissue, not by
hydrodynamics. The tissue itself is elastic on timescales of several
minutes~\cite{doubrovinski2017measurement}, comparable to our windows, and it
stiffens markedly during development~\cite{kiehart2000multiple}.
Viscoelasticity and confinement can make passive motion antipersistent but
cannot generate persistence, so the values $H>1/2$ we measure reflect the
activity of the cells, and any contribution from the medium can only bias $H$
downwards. This bias complicates the interpretation of the lower persistence
of cells in \textit{LanB1}-deficient embryos, whose surrounding matrix is
altered. Whether the cooling reflects decreasing activity or increasing
friction can be decided by three measurements: the dependence of $H$ on lag
time, which would reveal a relaxation time of the medium within our range of
lag times; the fluctuations of passive landmarks in the same embryos; and the
Kramers scaling of dwell times, $\ln\tau_{\mathrm{dwell}}\propto1/D_{\mathrm{eff}}$,
which annealing predicts and a change in friction alone does not.

The annealing picture makes further predictions that do not depend on the
value of $\gamma$. First, the cooling should be reversible. A wound, which
recruits hemocytes rapidly~\cite{stramer2005live}, should reheat the
population: annealing predicts a global rise in $D_{\mathrm{eff}}$, whereas
chemotaxis towards the wound should appear mainly as a local increase in $H$
directed at the wound. Second, if the cooling is set by the cells, direct
microrheology of the hemocoel, using magnetic beads or ferrofluid
droplets~\cite{doubrovinski2017measurement} at the stages imaged here, should
show a friction that changes much less than $D_{\mathrm{eff}}$. Third, an
actin or motility mutant with a normal matrix should change the temperature
but not the memory, which would separate a defect in motility from a defect in
the matrix.

There are some caveats to our conclusions. The six movies for each embryo were
recorded consecutively, and some changes in both parameters coincide with the
boundaries between movies. We have therefore based our conclusions only on
features that persist across several movies. The estimator assumes fractional
Brownian motion with a static localization error within each window. For other
processes, such as continuous-time random walks, L\'evy walks or diffusing
diffusivity, it returns the parameters of the closest fractional Brownian
description. Fisher orthogonality of $D_{\mathrm{eff}}$ holds for an efficient
estimator and is approximate for twenty-frame windows and for a Bayesian
estimator subject to prior shrinkage. Because the network does not report the
covariance between $H$ and $\log D$, the uncertainty of $\log D_{\mathrm{eff}}$
can only be bracketed. We used the upper bound, which makes our estimates of
heterogeneity conservative. $D_{\mathrm{eff}}$ depends on the frame interval
and window length whenever $H\neq1/2$, so comparisons with experiments at
other frame rates require converting back to $D$, which reintroduces the
uncertainty in $H$. Only the \textit{LanB1} deficiency was followed through
the patrolling phase; the collagen recordings cover the second and third hours
of imaging, and the comparison rests on four embryos.

The combination used here for the Cerberus neural network, calibrated local
estimates in coordinates whose errors do not mix, applies to any heterogeneous
anomalous transport recorded in short windows: endosomes and other
intracellular cargo, bacterial motility, immune cells in tissues, and tracers
in soft materials. In each case the generalized diffusion coefficient confuses
amplitude with memory, and $D_{\mathrm{eff}}$ separates them at no extra cost.
For hemocytes, it turns a qualitative change of behaviour during development
into a trajectory through the plane of temperature and memory. Cells in
control, \textit{LanB1}-deficient and collagen-IV-deficient embryos follow
different paths through this plane, and the paths can be compared
quantitatively. Robustness is a crucial defining principle in systems biology
for biologically relevant processes~\cite{alon2019introduction}. The
\emph{annealed search} strategy is expected to be robust to energy landscapes
with a wide variety of energy magnitudes. In the future it would be interesting
to explore the annealed search process with respect to adaptation and
fold-change detection, two important paradigms in systems biology that help
determine robustness.


\section*{Conclusion}

Hemocytes in the \textit{Drosophila} embryo switch within a few hours from
colonising the embryo to patrolling its tissues. We have described this
transition with two quantities: an effective diffusivity $D_{\mathrm{eff}}$,
which measures how strongly a cell moves, and the Hurst exponent $H$, which
measures how much it remembers its previous direction. Two tools made this
possible. A physics-informed neural network with a Cerberus architecture
estimates both $H$ and the generalized diffusion coefficient $D$ from twenty-frame
windows, each with its own uncertainty. The reparameterisation 
$D_{\mathrm{eff}}$ removes the degeneracy between amplitude and memory that is inherent to anomalous diffusion.
$D_{\mathrm{eff}}$ has units of $\mu\mathrm{m}^2\,\mathrm{s}^{-1}$ for every
$H$, equals $D$ for Brownian motion, and its error is asymptotically
uncorrelated with that of $H$. At the hemocyte frame interval a single window
fixes $D_{\mathrm{eff}}$ to about 20\%, while the errors of $H$ and $D$ are
strongly anticorrelated. Fractional Brownian motion with the inferred
parameters reproduces the measured frame-to-frame speed throughout
development, and the slowdown of the cells is fully accounted for by
$D_{\mathrm{eff}}$ and $H$.

In these coordinates hemocyte motility follows a clear course. During the first
hour of imaging the cells warm and become more persistent. Thereafter they
cool. In the period covered by recordings of all three genotypes, the initial
cooling is indistinguishable between control, \textit{LanB1}-deficient and collagen-deficient 
embryos. Later the genotypes diverge. Control hemocytes cool about threefold
while remaining clearly persistent ($H\approx0.63$). Hemocytes in \textit{LanB1}-deficient embryos are
hotter at first but cool about sixfold and approach Brownian motion. In control
embryos speed and persistence stay coupled throughout, so that fast cells keep
their direction even late in development. In \textit{LanB1}-deficient embryos the coupling
weakens steadily and vanishes in the last hour, when fast cells move without
directional memory. The spread between cells exceeds the estimation error in
$D_{\mathrm{eff}}$ at all times, and in $H$ for most of development, so the
embryo contains a genuinely heterogeneous population. The late differences between genotypes rest on one continuous recording per genotype and should therefore be interpreted with caution, although they are consistent across more than a thousand track segments per movie (Fig. \ref{fig:Hspeed}). The collagen mutant was recorded for the first three hours only, and its behaviour during the patrolling phase is not addressed here.

We interpret the cooling as an \emph{annealed search}: a high early
temperature disperses the population, and a falling temperature lets the cells
settle at favourable sites and patrol around them. In this picture the basement
membrane, through laminin, converts speed into persistence and so sustains the
directional memory of patrolling cells. The interpretation assumes that the
friction experienced by the cells does not rise during development. It makes
testable predictions: passive landmarks in the same embryos should not cool,
dwell times at favourable sites should follow Kramers' scaling, and wounding
should reheat the population.

More generally, the separation of amplitude from memory applies to all
heterogeneous anomalous transport. Wherever changes along a trajectory are to
be interpreted, a per-window uncertainty and a coordinate that does not mix the
two parameters are needed to tell genuine changes from artefacts of the
estimator.

\bibliography{sample}

\section*{Acknowledgements}

NK acknowledges useful discussions with Daniel Han. The authors used Claude Opus 4.8 to assist with reviewing and developing parts of the code, to assist with manuscript revision. The authors reviewed and edited all AI assisted content and take full responsibility for the content of the publication.

\section*{Author contributions statement}

NK: Conceptualization, Resources, Software, Formal analysis, Validation, Investigation, Visualization, Methodology, Project administration, Writing - original draft and editing; TM: Data curation, Validation, Investigation, Methodology, Writing - editing. TW: Formal analysis, Validation, Investigation, Methodology, Writing - editing.





\setcounter{figure}{0}
\renewcommand{\figurename}{Fig.}
\renewcommand{\thefigure}{S\arabic{figure}}

\setcounter{equation}{0} 
\renewcommand{\theequation}{S\arabic{equation}}


\section*{Supplementary Material}

\subsection*{S1. Network architecture, training, error calibration}

\subsubsection*{The fractional Gaussian noise correlation function}

The network operates on displacements rather than positions, because the
increments of FBM are stationary while the positions are not. For a window of
$N$ points sampled at an interval $\Delta t$, we write the displacements on axis
$c$ as $\delta^{(c)}_i = y^{(c)}_{i+1} - y^{(c)}_i$, with $i = 1,\ldots,n$ and
$n = N - 1$. These displacements are the fractional Gaussian noise (fGn), and
their correlation function is
\begin{equation}
  \varrho(k) = \langle \delta_i \delta_{i+k}\rangle
   = D\,\Delta t^{2H}\left(|k+1|^{2H} - 2|k|^{2H} + |k-1|^{2H}\right),
  \label{eq:fgn}
\end{equation}
with $\varrho(0) = 2D\Delta t^{2H}$, in the convention
$\langle \Delta x^2(t)\rangle = 2Dt^{2H}$ per axis used throughout this work.
Equation~\ref{eq:fgn} contains all of the information the network can use.
Persistent motion ($H>0.5$) gives positive correlations that decay slowly as
$k^{2H-2}$; anti-persistent motion ($H<0.5$) gives negative correlations that
sum to zero; and at $H = 0.5$ the correlations vanish and the displacements
are uncorrelated. It is the shape of Equation~\ref{eq:fgn}, and not the size
of the displacements, that determines $H$.

Experimental trajectories are recorded with a static localization error. We
write the measured positions as
\begin{equation}
  \mathbf{y}_k = \mathbf{X}(t_k) + \boldsymbol{\varepsilon}_k, \qquad
  \boldsymbol{\varepsilon}_k \sim
  \mathcal{N}\left(\mathbf{0},\sigma_{\mathrm{loc}}^2\,\mathbb{I}_3\right),
  \label{eq:obs}
\end{equation}
where $\boldsymbol{\varepsilon}_k$ has zero mean and is independent between
frames and between axes, and $\sigma_{\mathrm{loc}}$ is the standard deviation
of the measured position of an immobile particle. Differencing
Equation~\ref{eq:obs} shows that each displacement carries the error of two
frames, which raises its variance by $2\sigma_{\mathrm{loc}}^2$, while
consecutive displacements share one frame with opposite sign, which gives a
covariance of $-\sigma_{\mathrm{loc}}^2$ at lag one and zero at longer lags.
The covariance matrix of the observed displacements is therefore
\begin{equation}
  \Sigma_{ij} = \varrho(|i-j|)
    + \sigma_{\mathrm{loc}}^2\left(2\delta^{\mathrm{K}}_{ij}
      - \delta^{\mathrm{K}}_{|i-j|,1}\right),
  \label{eq:Sigma}
\end{equation}
where $\delta^{\mathrm{K}}$ is the Kronecker delta, written with a superscript
to distinguish it from the displacements $\delta^{(c)}_i$. The second term of Equation~\ref{eq:Sigma} is the zero-motion-blur limit of the
displacement covariance derived by Berglund \cite{berglund2010statistics} for
camera-based tracking of a Brownian particle, and the corresponding result for
FBM, in which $\varrho(|i-j|)$ replaces the Brownian term, was given by Weber
\emph{et al.} \cite{Weber2012}. At the level of the MSD the same effect appears
as the additive offset $2\sigma_{\mathrm{loc}}^2$ identified by Martin \emph{et
al.} \cite{martin2002apparent} and refined by Savin and Doyle \cite{Savin2005}. That the
lag-one term carries information as well as bias was shown by Vestergaard
\emph{et al.} \cite{Vestergaard2014}, whose covariance-based estimator (CVE)
recovers both the diffusion coefficient and the localization error of a
Brownian particle from the variance and the lag-one covariance of the
displacements alone. The CVE is explicit, unbiased and regression-free, and in
experimentally relevant parameter ranges it outperforms not only estimators
built from the measured MSD but also the maximum-likelihood estimator. Our
network may be considered as the FBM counterpart of that idea. Rather than fitting a
power law to an MSD curve, we estimate $H$, $D$ and $\sigma_{\mathrm{loc}}$
from the correlation structure of the displacements itself, which is why the
localization error appears as an output rather than as a nuisance to be
corrected for afterwards.

We assume throughout that the camera exposure is short compared with the frame
interval. For continuous illumination the particle moves during the exposure,
and its average position rather than its instantaneous position is
recorded \cite{Savin2005,berglund2010statistics}. This motion blur adds a positive
contribution to the lag-one term, partially cancelling the localization error,
and it is quantified by a single motion-blur coefficient determined by the
illumination profile \cite{berglund2010statistics,Vestergaard2014}. It therefore enters
Equation~\ref{eq:Sigma} as a known constant and could be included without any
change to the network.

That static localization error induces apparent correlations in measured
displacements is well established for Brownian
motion \cite{Savin2005,Michalet2010,MichaletBerglund2012,Weber2012}, and the
consequences for anomalous exponent estimation were analysed by Burnecki \emph{et
al.} \cite{burnecki2015estimating} and Lanoiselée \emph{et al.}  \cite{lanoiselee2018optimal}. The
negative lag-one term is what makes the estimation problem hard. An
anti-persistent FBM produces a negative correlation at lag one for exactly the
same reason that measurement noise does, so the two are difficult to separate
in a short window, and the difficulty is worst at small $H$ where the physical
correlation is itself negative. We define the signal-to-noise ratio as
$\mathrm{SNR} = \sqrt{2D\Delta t^{2H}}/\sigma_{\mathrm{loc}}$, the ratio of the
standard deviation of the noise-free displacements to the standard deviation of
the localization error of a single position, in agreement with our earlier
definition.

\subsubsection*{Feature engineering}

Which features from the data should be given to the Cerberus neural network for training? As we found previously with the
tandem NNs \cite{korabel2025deep}, the choice of feature engineering is more important
than the choice of architecture, and we describe our criteria here. As a first
criterion, the features must be invariant under a change of units, so that one
trained network can be applied to data recorded in $\mu$m or in pixels. As a
second criterion, they must be invariant under a change of the frame rate, so
that the same network can be used at any $\Delta t$. As a third criterion, they
must not require the range of $D$ to be fixed in advance, since in the tandem
long short-term memory (LSTM) neural networks the range of $D$ had to be customized for each application \cite{korabel2025deep}.

All three criteria are met by normalizing the displacements by their own root
mean square,
\begin{equation}
  s^2 = \frac{1}{3n}\sum_{c=1}^{3}\sum_{i=1}^{n}
        \left(\delta^{(c)}_i\right)^2,
  \label{eq:s2}
\end{equation}
which is the time-averaged MSD of the window at its shortest lag, averaged
over the three axes. The input layer is loaded with the $3n$ normalized
displacements $\delta^{(c)}_i/s$. The network is given neither $\Delta t$ nor
the units of the coordinates. Since FBM is self-similar, the normalized
displacements have the same statistics for any length scale and any frame
rate, and the network is left to learn the shape of Equation~\ref{eq:fgn}
alone.

The scale was removed by Equation~\ref{eq:s2} and we restore it afterwards.
From Equation~\ref{eq:Sigma}, $\langle s^2\rangle = 2D\Delta t^{2H} +
2\sigma_{\mathrm{loc}}^2$, and inverting the first term gives
\begin{equation}
  \log D = r + \log s^2 - 2H\log \Delta t - \log 2,
  \label{eq:logD}
\end{equation}
where the last three terms are the plug-in MSD estimate and $r$ is a
correction predicted by the network. The correction absorbs what the plug-in
estimate omits: the finite-$n$ bias of $s^2$, the $2\sigma_{\mathrm{loc}}^2$
offset, and the correlation between the errors in $s^2$ and in the predicted
$H$. The correction is of order unity in decades, while $D$ itself varies over several
decades, so the network learns a small quantity rather than the magnitude of
$D$. This is what allows one trained network to be used over the whole range
of $D$ without retraining. We note that Equation~\ref{eq:logD} also shows why
$H$ and $D$ must be estimated together: an error in $H$ enters $\log D$
multiplied by $2\log \Delta t$.

Two further inputs are supplied. The first is a flag which states whether the localization error
($\sigma_{\mathrm{loc}}$) has been measured independently, for example from
movies of immobilized beads. The second is $\log (\sigma_{\mathrm{loc}}/s)$, which is
used when the flag is set and replaced by a fixed value otherwise. The size of
the input layer is therefore $3(N-1)+2$.

Because the network is trained with a squared error, its outputs approximate
posterior means~\cite{han2026maximum} and are shrunk towards the centre of the
training distribution wherever the data carry little information: this pulls
$\hat H$ inwards near the edges of its range, lowering its coverage there, and
leaves the estimate of $\sigma_{\mathrm{loc}}$ close to its prior mean.

\subsubsection*{Cerberus neural network architecture}

The network consists of a shared trunk and three output heads; a \emph{Cerberus architecture}. The trunk is a
feedforward network with four layers of 256, 256, 128 and 64 units and SiLU
activation functions. Each head is a two-layer network acting on the 64
outputs of the trunk. We used a common trunk because $H$, $D$ and
$\sigma_{\mathrm{loc}}$ are all read from the same correlation function, and
separate heads because they are read from different parts of it.

The first head predicts the Hurst exponent in the output layer with the
sigmoid activation function, rescaled so that $0.02 < H < 0.98$, that is,
$0.04 < \alpha < 1.96$. The second head predicts the correction $r$ of
Equation~\ref{eq:logD} with the hyperbolic tangent activation function. 
The third head predicts $\sigma_{\mathrm{loc}}$ as a fraction of the
root mean square displacement, bounded between $10^{-2}$ and $1$. The lower
bound was chosen because a localization error below one per cent of the
displacements changes the variance by less than $0.02\%$ and cannot be
recovered from a short window.  $\sigma_{\mathrm{loc}}$ should, where possible, be measured
independently, for example from immobilised beads or fixed cells imaged under
the same conditions, or from whole trajectories pooled over a movie; a value
measured in this way is supplied to the network as an input and replaces the
output of this head.

The first two heads also predict the standard deviations, $\sigma_H$ and
$\sigma_{\log D}$. These are the uncertainties which we report, and which
are combined when the window is slid along the trajectory. They are the
predictive spread of a single trained network.

\subsubsection*{Loss function}

The network was trained with a loss function which combines the exact
likelihood of the FBM model with the supervised error on the simulated
parameters,
\begin{equation}
  L = w(e)\,L_{\mathrm{phys}} + L_{H} + L_{\log D}
      + \lambda_\sigma L_\sigma ,
  \label{eq:total}
\end{equation}
and we describe each of the four terms in turn.

Since the displacements are Gaussian with the covariance matrix of
Equation~\ref{eq:Sigma}, and the three axes are independent and share
$\Sigma$, the likelihood term is
\begin{equation}
  L_{\mathrm{phys}} = \frac{1}{3n}\left[\frac{1}{2}\sum_{c=1}^{3}
    \boldsymbol{\delta}^{(c)\mathsf{T}}\Sigma^{-1}\boldsymbol{\delta}^{(c)}
    + \frac{3}{2}\ln\det\Sigma + \frac{3n}{2}\ln 2\pi\right],
  \label{eq:phys}
\end{equation}
evaluated at the values of $H$, $D$ and $\sigma_{\mathrm{loc}}$ predicted by
the network and computed with a Cholesky factorization of $\Sigma$. 
Here $\boldsymbol{\delta}^{(c)}$ is the vector of the $n = T-1$ observed displacements on axis $c$, $\Sigma$ is the $n\times n$ covariance matrix of
Equation~\ref{eq:Sigma} built from the predicted $H$, $D$ and
$\sigma_{\mathrm{loc}}$ and shared by the three axes, $\det\Sigma$ is its
determinant, and the last term is a normalisation constant. The
quadratic form measures how well the predicted correlation function accounts
for the observed displacements, and is small when $\Sigma$ whitens the data.
The term $\ln\det\Sigma$ is the normalization of the Gaussian and prevents
$\Sigma$ from being made large to reduce the first term; together they fix the
scale of $D$ as well as the shape set by $H$. The constant carries no gradient
and is retained only so that the printed value is a likelihood. We divide by
$3n$ so that the term is a per-displacement quantity and does not change its
weight relative to the others when the window size is changed. The logarithms
here are natural, since they belong to the Gaussian normalization and are not
a reporting convention. Equation~\ref{eq:phys} does not use the ground truth,
so it constrains the network on experimental windows as well as on simulated
ones, and it excludes combinations of the three parameters which are
inconsistent with the measured displacements.

The point estimates and their uncertainties were trained together with the
heteroscedastic Gaussian loss
\begin{equation}
  L_\theta = \left\langle \frac{(\hat\theta - \theta)^2}{2\sigma_\theta^2}
    + \ln\sigma_\theta\right\rangle, \qquad
  \theta = H,\ \log D,
  \label{eq:hetnll}
\end{equation}
one for each estimated parameter. These are the only terms that use the
simulated ground truth. The squared error divided by the predicted variance
trains the point estimate, and $\ln\sigma_\theta$ prevents the network from
reducing it by reporting an arbitrarily large uncertainty; the minimum with
respect to $\sigma_\theta$ lies at the root mean squared error of the
prediction for that window. This is why $\sigma_H$ and $\sigma_{\log D}$
are calibrated by the training itself and require no subsequent rescaling. 
The fourth term in equation S6, $L_\sigma$, is a squared error on
$\log(\sigma_{\mathrm{loc}}/s)$ against the clipped target, and it trains the third head. 
The likelihood is sharply peaked in $H$ and, without
this, it dominates the gradient before the trunk has learned anything about the correlation function. All four terms are averaged over the batch.

\subsubsection*{Generation of training and validation datasets}

Synthetic FBM trajectories with constant Hurst exponents $H$ and generalized diffusion
coefficients $D$ were used to train the network, and a new batch was
simulated at every epoch, so that no window was used twice. For each
trajectory we drew $H$ uniformly in the range $(0.04, 0.96)$ and $\log D$ uniformly
in $(-5, 1)$, that is $D \in (10^{-5}, 10^{1})$. The displacements were generated directly from the correlation function rather
than from the Langevin equation. We factorized the covariance matrix of
Equation~\ref{eq:Sigma} with $\sigma_{\mathrm{loc}} = 0$ and multiplied it by a
vector of standard normal variates, obtained the positions by cumulative
summation, and then added Gaussian noise with the standard deviation
$\sigma_{\mathrm{loc}} = \sqrt{2D\Delta t^{2H}}/\mathrm{SNR}$ to represent the
static measurement error. The simulator and the likelihood of
Equation~\ref{eq:phys} therefore use the same covariance matrix.

The value of $\sigma_{\mathrm{loc}}$ was supplied to the network for half of
the trajectories in each batch. A single trained network can therefore be used
both when the localization error has been calibrated and when it has not, and
the difference between the two shows how much of the uncertainty in $H$ is
caused by the localization error.

The network was trained for $10^4$ epochs with batches of 512 trajectories,
using AdamW with a learning rate of $3\times10^{-4}$, a weight decay of
$10^{-4}$ and gradient clipping. Every 100 epochs we evaluated the network on
eight independently simulated validation batches, halved the learning rate
when the validation loss reached a plateau, and saved the network when the
validation loss improved. In addition to the MAE of $H$ and $\log D$, we
recorded the fraction of windows for which the true value fell inside
$\hat\theta \pm 1.96\sigma_\theta$. This coverage is as important as the MAE:
if the predicted uncertainties are too small, the variations of $H$ and $D$
along a trajectory cannot be distinguished from the errors of the estimate.

\subsubsection*{Calibration of the uncertainties}

Should the predicted uncertainties be rescaled after training? We chose not to apply any calibration, and we describe our reasons here. The standard
deviations $\sigma_H$ and $\sigma_{\log D}$ reported by the network are the
raw outputs of the two heads. No variance inflation
factor and no post-hoc regression of the predicted uncertainty against the
observed error were used at any stage.

The first reason is that the uncertainties are calibrated during training and
not afterwards. The loss function described by Equation~\ref{eq:hetnll} is minimized when
$\sigma_\theta^2$ equals the mean squared error of the prediction for that
window, so the network is trained to report the error it actually makes. A
network trained with a plain squared error, as in our previous work, produces
no uncertainty at all, and any interval must then be attached to it afterwards
from the ensemble statistics. Here the interval is a per-window quantity, and it
is learned together with the point estimate. The distinction matters for
heterogeneous trajectories: the error of the estimate depends strongly on $H$
itself, being largest at small $H$ where the localization error imitates
anti-persistence, so a single ensemble-averaged error bar would be too large in
some parts of a trajectory and too small in others.

The second reason is that a calibration factor cannot be transferred. Any
factor would have to be fitted on simulated windows drawn from the training
distribution, that is, on FBM with a particular range
of $D$, of $\Delta t$, and of the SNR. Experimental trajectories are not drawn
from that distribution. Applying a factor fitted on synthetic FBM to endosomes
or to tracers in a gel would replace an uncertainty which is at least
internally consistent with the model by one which has been adjusted using data
of a different kind. We prefer to report the uncertainty which follows from the
FBM likelihood and to test it separately.

The third reason is that the coverage of the uncertainties is itself a useful
diagnostic, and rescaling would remove it. In our previous work we compared the
NN predictions with the tMSD analysis, which assumes no anomalous diffusion
model, to check that FBM was an appropriate description of the data.
The coverage plays the same role here. If the true value falls inside
$\hat\theta\pm1.96\sigma_\theta$ for close to $95\%$ of windows, the FBM model
with a static localization error describes the data; if the coverage is much
smaller, either the model is wrong, or the trajectories lie outside the range on
which the network was trained. A calibration factor chosen to force the
coverage to $95\%$ would conceal exactly the discrepancy which we want to
detect. We therefore compute the coverage independently, with a separate
analysis script, and report it alongside the MAE rather than using it to adjust
the network.

We employ a Physics-informed Neural Network (PINN) with a \emph{Cerberus architecture} that learns the spatio-temporal features directly from the displacements of trajectory data. A schematic diagram is illustrated in Fig.~\ref{fig:fig0}. We use the PINN to estimate the Hurst exponent $H$, related to the anomalous exponent of the MSD by
$H = \alpha/2$, $\langle r^2(t)\rangle = 2dDt^{\alpha}$ (where $d$ is the number of spatial dimensions, which we have considered $d=3$ in the following), the generalized diffusion coefficient $D$ and the static localization error $\sigma_{\mathrm{loc}}$ from short three-dimensional windows of a single trajectory. $\sigma_{\mathrm{loc}}$ is well defined in the literature \cite{Savin2005,backlund2015chromosomal} and represents the finite time-independent uncertainty (static localization error) on measuring the position of particles in an optical microscope. 
In contrast to our previous tandem approach \cite{korabel2025deep}, where the estimate of $H$ is passed from H-LSTM to a second D-LSTM network, here a single PINN estimates $H$ and $\log D$ simultaneously and returns uncertainties in the form of standard deviations, $\sigma_H$ and $\sigma_{\log D}$.

To assess whether estimators of anomalous exponent and diffusion coefficient are trustworthy, we simulated a set of  FBM trajectories of duration $T$ with constant Hurst exponents $0<H<1$ and generalized diffusion coefficients in the range of $\log D \in (-5, 1)$. The predictive capabilities of neural networks depend heavily on the available information. For example, the duration of the trajectories ($T$) is important as well as the amount of measurement noise. Since our ultimate goal is to quantify heterogeneous trajectories whose properties change over time using fixed $T$ trajectories combined with a rolling window, we start with a minimum duration of $T = 20$ time points and with the small Gaussian measurement noise added to trajectories with a fixed signal-to-noise ratio (SNR). Together with the $0.02$ second time increment used to generate FBM trajectories, this corresponds to $0.4$ second trajectories. We will later show how the NN predictions, their error, and calibration depend on $T$ and on the signal-to-noise ratio.

The performance of the PINN to predict the Hurst exponent for short $20$ time-point FBM trajectories is illustrated in Fig.~\ref{fig:figS1}(A). The dense concentration of points along the identity line indicates that the network successfully captures the relationship between particle trajectories and the underlying Hurst exponent across the entire interval $0<H<1$. The colour scale, representing the absolute prediction error $|H_{\mathrm{pred}}-H_{\mathrm{theor}}|$, reveals that most trajectories are reconstructed with only small deviations from the true value. 
\begin{figure}[h!]
  \centering  \includegraphics[width=0.95\textwidth]{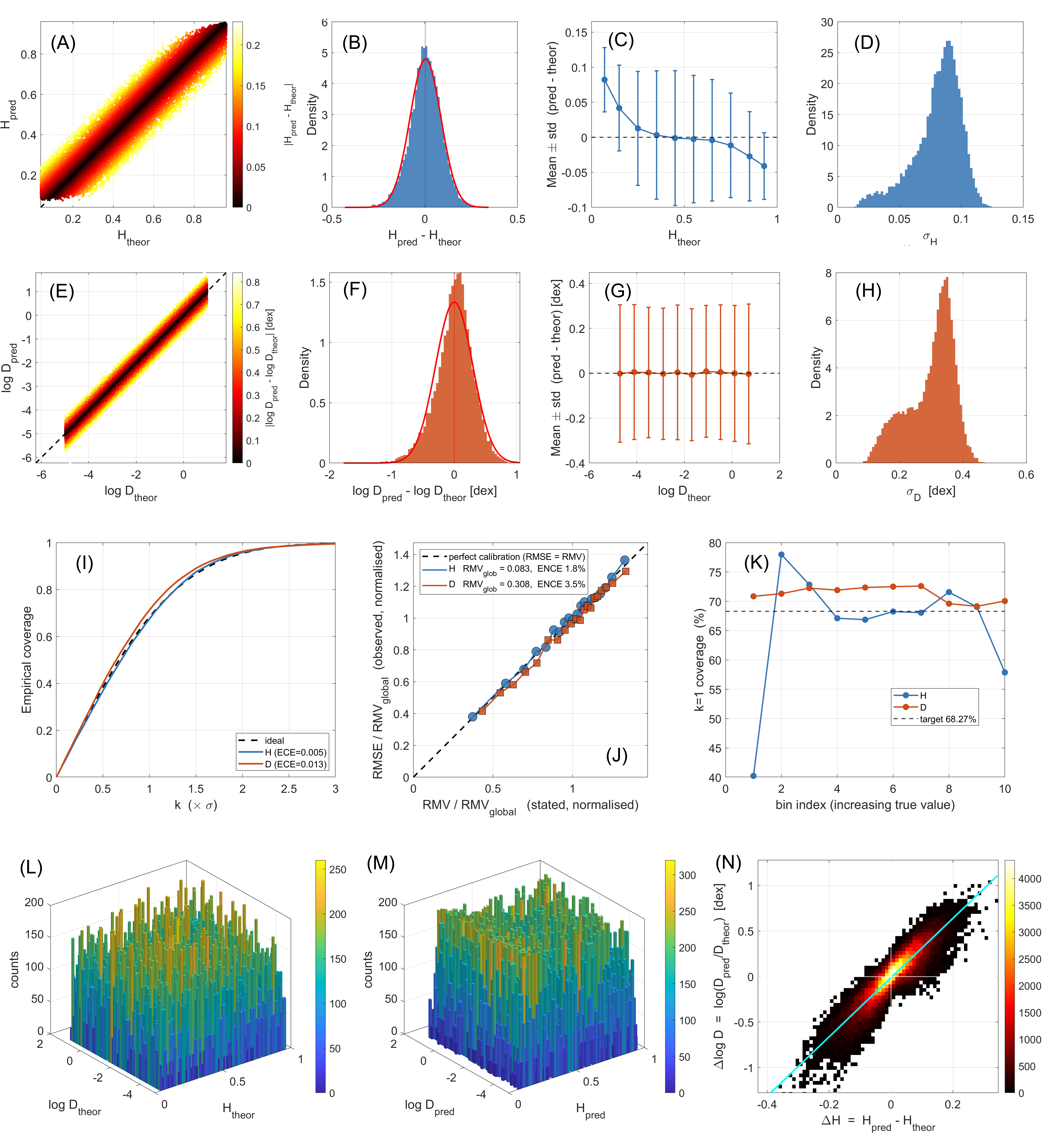}
  \caption{Performance of the neural-network estimator for predicting the Hurst exponent $H$ and generalised diffusion coefficient $D$ from very short simulated FBM trajectories of 20 time points (19 displacements with $\Delta t = 0.02$~s per frame) with constant $H$, constant $D \in (10^{-5}, 1)$, motion blur (dynamical error) and static measurement error with SNR=20. \textbf{(A, E)} Scatter plot of the predicted Hurst exponent,
$H_{\mathrm{pred}}$ and generalised diffusion coefficient $\log D_{pred}$, against the ground-truth values, $H_{\mathrm{theor}}$ and $D_{\mathrm{theor}}$. The black diagonal denotes the ideal predictions. The colour scales represent the absolute prediction errors. 
\textbf{(B, F)} Distributions of the signed prediction errors.The histograms are compared with a zero-mean Gaussian having the same
variance (red solid lines). \textbf{(C, D)} Mean prediction bias stratified by the true values. For each interval of values, the marker denotes the mean prediction error, and the error bars represent one standard deviation. \textbf{(D, H)} Distributions of predicted standard deviations  $\sigma_H$ and $\sigma_D$. Assessment of uncertainty calibration and reliability for variables $H$ and $D$ \textbf{(I, J, K)} (continued on next page). 
}
  \label{fig:figS1}
\end{figure}
\begin{figure}[htbp]
    \ContinuedFloat 
    \captionsetup{list=no} 
    \caption[]{(Continued) \textbf{(I)} Coverage reliability diagram demonstrating empirical coverage against the standard deviation multiplier, $k (\times \sigma)$. \textbf{(J)} Reliability plot comparing normalized  RMSE to RMV to evaluate variance calibration. \textbf{(K)} Conditional coverage for $k=1$, evaluated across 10 bins 
    compared against the ideal target of 68.27\%. \textbf{(L)} Histogram of the true parameter space. \textbf{(M)} Corresponding 3D histogram of the network's estimated parameters. \textbf{(N)} Joint error distribution residuals. The cyan line highlights the principal axis of the strongly correlated error landscape.}
\end{figure}
Quantitatively, this visual impression is confirmed by the low Mean Absolute Error (MAE $=0.058$) and root-mean-square error (RMSE $=0.07$). The histogram of the signed residuals in Fig.~\ref{fig:figS1}(B) is centred very close to zero and is well approximated by a Gaussian distribution having the same mean and variance. 
This confirms that the estimator is essentially unbiased and that its errors are dominated by random fluctuations rather than systematic offsets. Fig.~\ref{fig:figS1}(C) shows a monotonic decrease of the mean prediction error as a function of the true Hurst exponent. For strongly subdiffusive trajectories, the network tends to slightly overestimate $H$, whereas for strongly superdiffusive trajectories it slightly underestimates it. Importantly, the magnitude of this systematic bias remains small throughout the entire range of $H$ and is considerably smaller than the intrinsic prediction uncertainty represented by the standard deviations within each bin. Fig.~\ref{fig:figS1}(D) shows the distribution of Hurst exponent error $\sigma_H$ reported by the PINN.

The performance of the PINN to predict the generalised diffusion coefficient for short $20$ time-point FBM trajectories is illustrated in Fig.~\ref{fig:figS1}(E). Because $D$ spans multiple orders of magnitude, the network evaluates the target in logarithmic space. The dense concentration of points along the identity line demonstrates the network's strong predictive capability across the entire evaluated range. Quantitatively, this is supported by the low Mean Absolute Error (MAE = 0.53), root-mean-square error (RMSE = 0.3), and the negligible mean bias (-0.02). The histogram of the signed residuals in Fig.~\ref{fig:figS1}(F), defined as $\log D_{\mathrm{pred}} - \log D_{\mathrm{theor}}$, is centred close to zero and is well approximated by a Gaussian distribution having the same mean and variance. This confirms that the estimator is essentially unbiased in logarithmic space, with errors dominated by random fluctuations rather than systematic offsets. In contrast to the Hurst exponent predictions, Fig.~\ref{fig:figS1}(G) shows no systematic bias in the mean prediction error as a function of the true diffusion coefficient. The mean error profile remains flat and close to zero across all bins, and the intrinsic prediction uncertainty remains remarkably stable regardless of the magnitude of $D$. The distribution of standard deviations of predicted diffusion coefficients is shown in Fig.~\ref{fig:figS1}(H).

The main question is how reliable and calibrated the network's outputs are. Fig.~\ref{fig:figS1}(I-K) demonstrates that both Hurst exponent $H$ and diffusion coefficient $D$ achieve excellent global calibration. By looking at the coverage in Fig.~\ref{fig:figS1}(I), which represents the empirical proportion of true observation values that fall within a predicted uncertainty interval defined by a standard deviation multiplier ($k$), we find that both variables closely track the ideal coverage reliability curve. Because their empirical coverage closely matches the ideal expectation, both variables exhibit exceptionally low Expected Calibration Error (ECE), which quantifies the average absolute difference between the observed empirical coverage and the theoretical confidence level across all evaluated intervals. $H$ achieves an ECE of 0.009, and $D$ achieves 0.008. This strong global performance is confirmed by Fig.~\ref{fig:figS1}(J), where both $H$ and $D$ points fall tightly along the perfect calibration diagonal. This alignment results in low Expected Normalized Calibration Error (ENCE) that assess variance calibration by measuring the average relative difference between the actual prediction error (RMSE) and the predicted uncertainty (RMV) across sorted bins, yielding 5.2\% for $H$ and 6.1\% for $D$. Fig.~\ref{fig:figS1}(K) shows that $D$ maintains stable, with consistent coverage near the 68.27\% target (representing an ideal $1\sigma$ interval) across all true value bins, while $H$ experiences severe under-coverage at the extreme ends of the distribution (bins 1 and 10). This indicates that $H$ is well-calibrated on average and its uncertainty estimates become overly confident and unreliable at the extreme boundaries of the data space, whereas $D$ remains robust throughout. 

The global fidelity of the PINN in reconstructing the joint parameter space for short $T=20$ time-point FBM trajectories is qualitatively illustrated by comparing the true parameter plane in Fig.~\ref{fig:figS1}(L) with the estimated parameter plane in Fig.~\ref{fig:figS1}(M). The macroscopic similarity between the two distributions confirms that the estimator faithfully reproduces the underlying bivariate parameter space without introducing severe large-scale distortions or collapsing the variance of the predictions. However, examining the joint error distribution in Fig.~\ref{fig:figS1}(N) reveals a critical dependency between the estimation errors of the two individual parameters. The 2D density plot of the residuals, $\Delta H$ and $\Delta \log D$, exhibits a pronounced diagonal elongation. This strong positive correlation, highlighted by the cyan trend line, indicates that the network's prediction errors are strongly coupled: an overestimation of the Hurst exponent systematically drives an overestimation of the generalised diffusion coefficient, and vice versa. This inherent degeneracy reflects a fundamental physical ambiguity in short-trajectory characterisation, wherein an increased estimated scaling exponent $H$ requires a corresponding compensatory increase in the diffusion amplitude $D$ to adequately map to the observed trajectory fluctuations.


\subsubsection*{PINN performance as a function of trajectory length and measurement noise}

\begin{figure}[htbp]
  \centering  \includegraphics[width=0.95\textwidth]{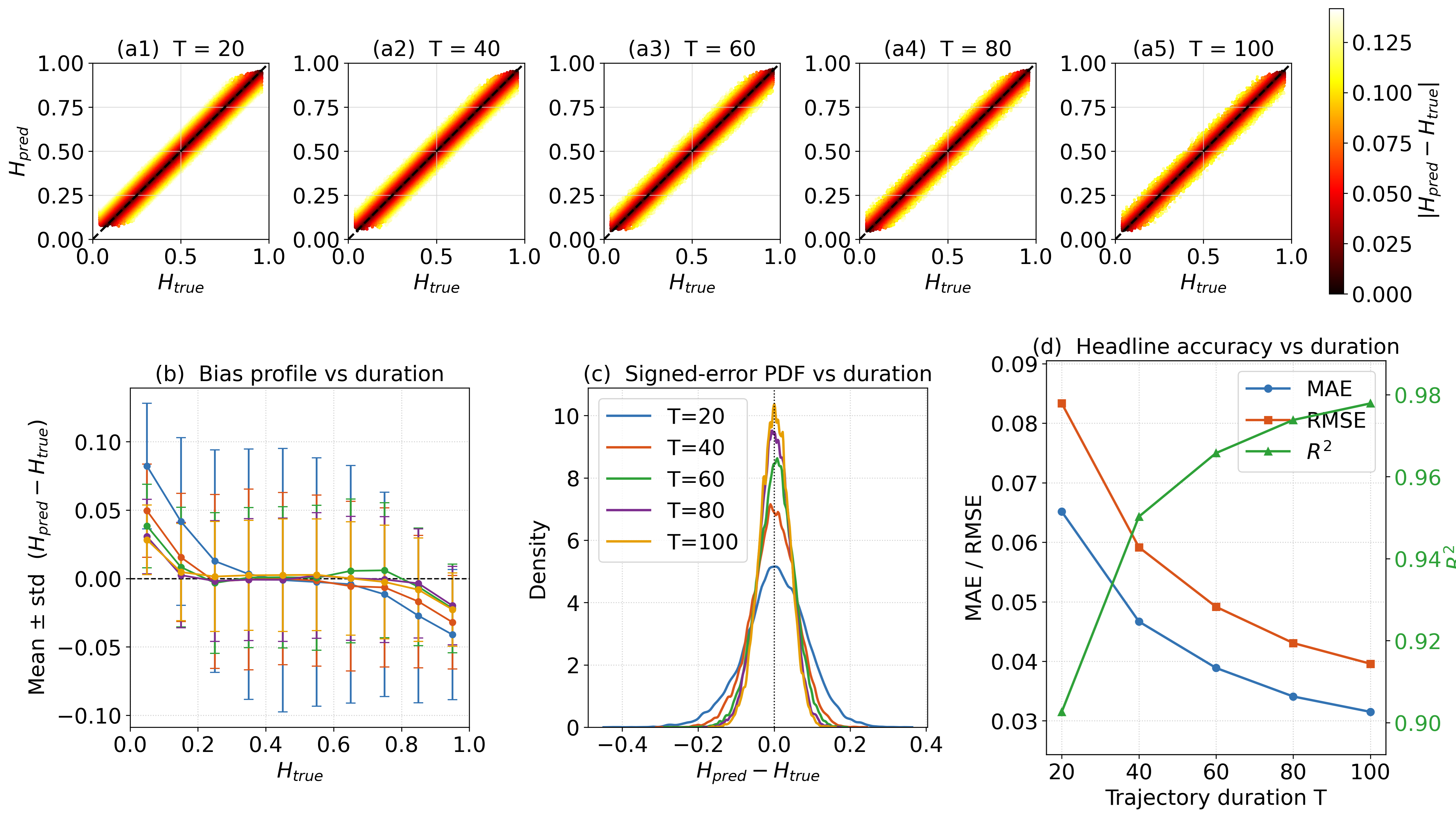}
  \caption{Performance of the PINN estimator for predicting the Hurst exponent $H$ from simulated FBM trajectories of varying durations $T = 20, 40, 60, 80,$ and $100$ time points ($T-1$ increments). Trajectories were simulated with $dt=0.02$, constant uniform $H\in (0.04, 0.96)$, constant uniform $\log D \in (-5, 1)$ and SNR=20.  (a1)–(a5) Density scatter plots comparing the predicted Hurst exponent ($H_{pred}$) against the theoretical ground truth ($H_{theor}$). The colormap indicates the absolute prediction error, $|H_{pred} - H_{theor}|$.  (b) Bias profile illustrating the mean signed error ($H_{pred} - H_{theor}$) as a function of the theoretical Hurst exponent across different trajectory durations. Error bars represent one standard deviation, highlighting the reduction in variance as $T$ increases.(c) Probability density functions (PDF) of the signed prediction errors. The distributions become progressively narrower and more sharply peaked at zero for longer trajectories, confirming improved model calibration. (d) Headline accuracy metrics plotted against trajectory duration $T$. Mean Absolute Error (MAE, blue circles) and RMSE (orange squares) correspond to the left y-axis, demonstrating a steady decay in error. The corresponding rise in $R^2$ (green triangles) is tracked on the right y-axis.  
}
  \label{fig:S1}
\end{figure}

\begin{figure}[htbp]
  \centering  \includegraphics[width=0.95\textwidth]{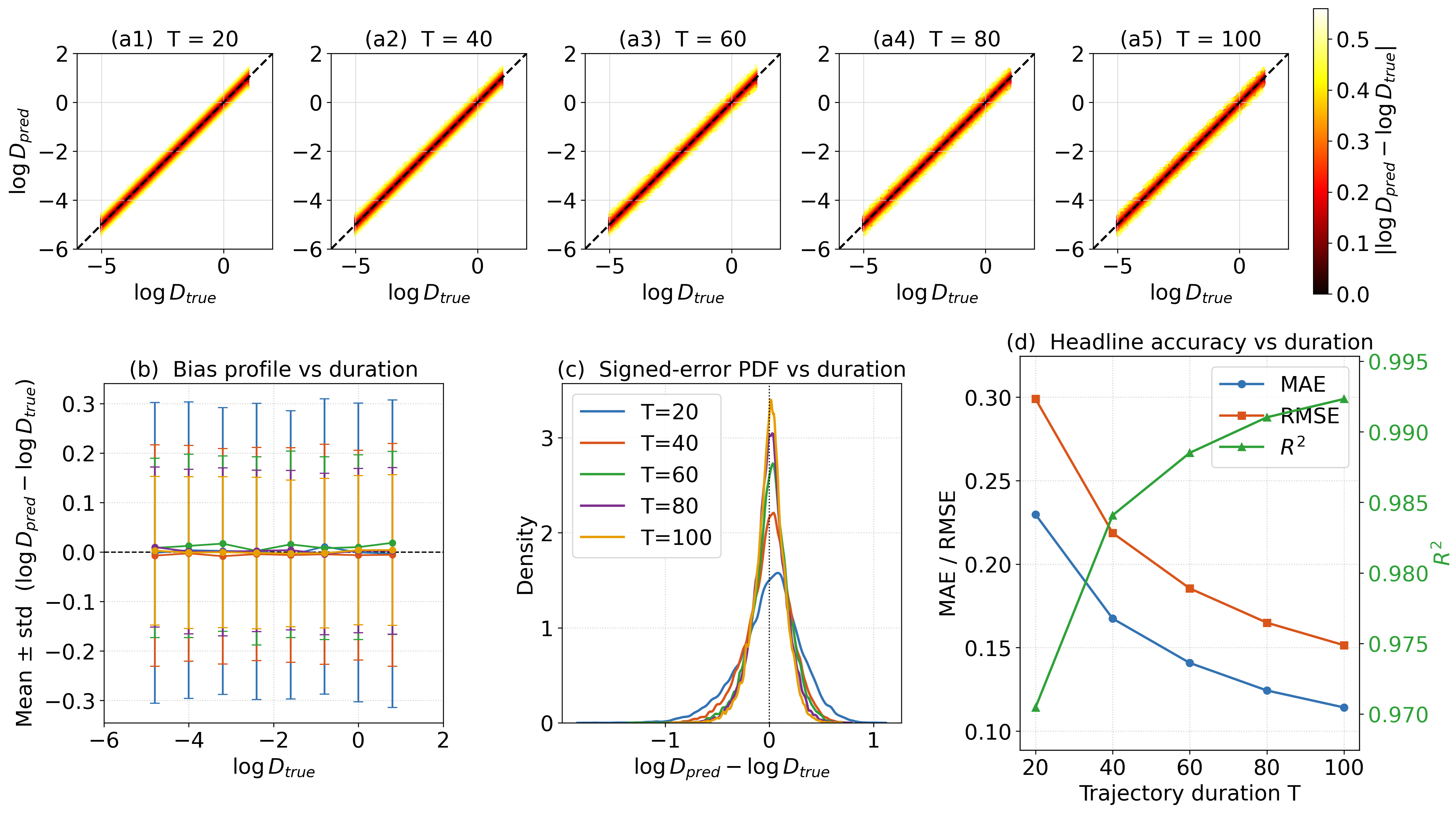}
  \caption{Same as in Fig. S2, but for generalised diffusion coefficient. 
}
  \label{fig:S2}
\end{figure}

The performance scaling of the PINN estimator with respect to trajectory length is shown in figures~\ref{fig:S1} and \ref{fig:S2}. It demonstrates a monotonic improvement in predictive reliability as the trajectory duration $T$ increases from 20 to 100 time points. While the model successfully captures the overall Hurst exponent trend at highly constrained lengths (e.g., $T=20$), the predictions naturally exhibit a wider variance around the theoretical ground truth. Extending the trajectory length rapidly resolves this dispersion, visibly tightening the prediction spread and elevating the coefficient of determination ($R^2$) to 0.98 by $T=100$. Furthermore, an analysis of the signed prediction errors confirms the model remains exceptionally well-calibrated; the error distributions center symmetrically at zero and sharpen significantly as more temporal data is introduced. Crucially, the estimator maintains a stable, near-zero bias profile across the entire functional range of $H$, indicating that the network generalizes effectively without severe boundary degradation, even when processing highly abbreviated  FBM sequences.

\begin{figure}[htbp]
  \centering  \includegraphics[width=0.95\textwidth]{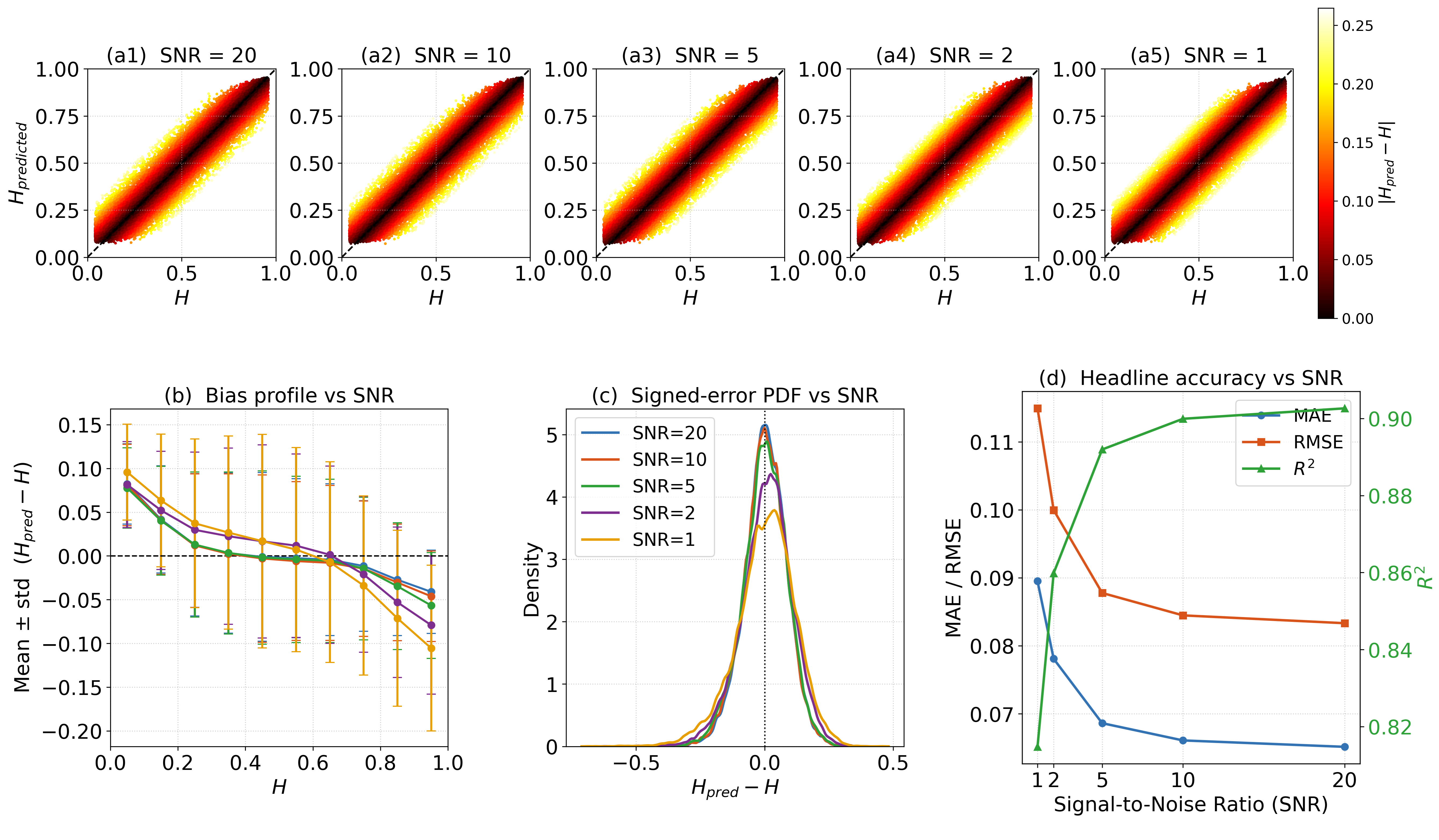}
  \caption{Performance of the PINN estimator for predicting the Hurst exponent $H$ from simulated FBM trajectories with different Signal-to-Noise Ratios (SNR). Trajectories of durations $T = 20$ time points ($19$ increments) were simulated with $dt=0.02$, constant uniform $H\in (0.04, 0.96)$, constant uniform $\log D \in (-5, 1)$ and SNR = 20, 10, 5, 2, and 1.  (a1)–(a5) Density scatter plots comparing the predicted Hurst exponent ($H_{pred}$) against the theoretical ground truth ($H_{theor}$). The SNR is given in the title of each panel. The colormap indicates the absolute prediction error, $|H_{pred} - H_{theor}|$.  (b) Bias profile illustrating the mean signed error ($H_{pred} - H_{theor}$) as a function of the theoretical Hurst exponent across different trajectory durations. Error bars represent one standard deviation, highlighting the reduction in variance as $T$ increases.(c) Probability density functions (PDF) of the signed prediction errors. The distributions become progressively narrower and more sharply peaked at zero for longer trajectories, confirming improved model calibration. (d) Headline accuracy metrics plotted against trajectory duration $T$. Mean Absolute Error (MAE, blue circles) and RMSE (orange squares) correspond to the left y-axis, demonstrating a steady decay in error. The corresponding rise in $R^2$ (green triangles) is tracked on the right y-axis.
}
  \label{fig:S3}
\end{figure}

The robustness of the PINN estimator to measurement noise is shown in Fig.~\ref{fig:S3}. Fixing the trajectory length to just $T=20$ time points, the estimator exhibits strong noise resilience in moderate noise regimes. Prediction accuracy, error variance, and overall model calibration remain remarkably stable down to an SNR of 5, maintaining a coefficient of determination ($R^2$) above 0.92. At extreme noise thresholds (SNR of 2 or lower), structural degradation in predictive performance becomes evident. In these highly corrupted signal environments, the model begins to systematically underestimate the true Hurst exponent, manifesting as a pronounced negative bias that becomes increasingly severe at higher theoretical $H$ values.

\begin{figure}[htbp]
  \centering  \includegraphics[width=0.95\textwidth]{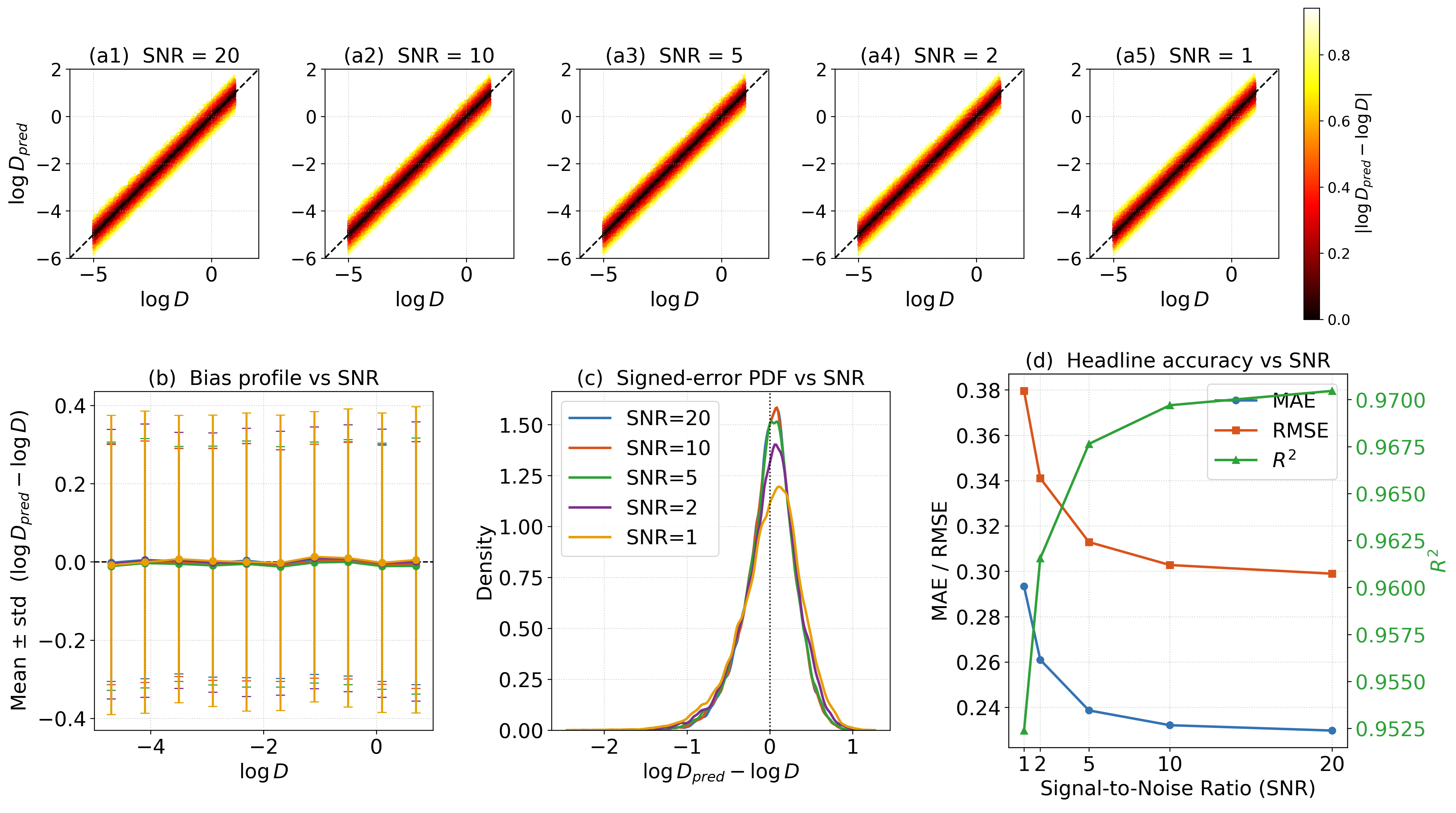}
  \caption{Same as in Fig. S4, but for the generalised diffusion coefficient. 
}
  \label{fig:S3N}
\end{figure}

\subsection*{S2. The effective diffusion coefficient $D_{\mathrm{eff}}$}
\label{sec:Deff}

As in the main text, $\log$ denotes $\log_{10}$, and logarithms of dimensional
quantities refer to their numerical values in \textmu m and s.

\subsubsection*{Why $D$ is a poor measure of amplitude}
For anomalous diffusion the mean squared displacement grows as
\begin{equation}
  \langle \mathbf{r}^2(t)\rangle = 2d\,D\,t^{\alpha},\qquad \alpha=2H
  \ \text{for fractional Brownian motion (FBM)}.
  \label{eq:msd}
\end{equation}
Two quantities are mixed in this expression: how strongly the particle moves,
and how much its motion remembers the past. The mixing shows up first in the
units. $D$ is measured in $\mathrm L^2\,\mathrm T^{-2H}$, so two cells with
different $H$ have diffusion coefficients that cannot be compared, much as a
velocity cannot be compared with an acceleration. Changing the time unit by a
factor $c$ shifts $\log D$ by $2H\log c$, a different amount for every cell.

The mixing shows up again in the errors. A short window measures the typical
size of a step very well, because every step samples it. It measures the
exponent much less well, because that requires following how displacements
grow over a narrow range of lags, about $1.3$ decades for twenty frames, with
the longest lags represented by only a few overlapping displacements. The data
therefore pin the MSD near the frame lag but leave its slope loose. An error in
$H$ rotates the fitted MSD, on log--log axes, about a pivot lag $\tau^*$, and
$D$ adjusts to keep the curve through the pinned point:
\begin{equation}
  \delta\log D \simeq -2\log\tau^*\,\delta H .
  \label{eq:ridge}
\end{equation}
The sign of this correlation depends only on whether $\tau^*$ is shorter or
longer than the unit of time. With $\tau^*\simeq0.5\,\Delta t$ (derived below),
it predicts slopes of $+4.0$\,dex per unit $H$ at $\Delta t=0.02$\,s and
$-1.8$\,dex at $\Delta t=17$\,s, as observed (Fig.~\ref{fig:degeneracy} in the main text). The correlation is a property of the measurement, not of the particle. What we need
is an amplitude that has the same units for every trajectory, that does not
inherit the error of $H$, and whose meaning does not depend on a particular
model.

\subsubsection*{Memory and fresh randomness}
The motion with memory can be pictured as a Langevin equation. Each step is the sum of a part predictable from the preceding steps and a fresh random kick that the past cannot anticipate. For a persistent cell the
predictable part is the tendency to keep going in the same direction; for an
antipersistent one it is the tendency to turn back. The kicks are the genuinely
new randomness of the motion, and their strength is the natural analogue of a
temperature: a thermometer should respond to fluctuations, not to a cell that
simply keeps moving straight.

To make this quantitative, consider $N$ consecutive steps of one coordinate,
$\Delta x_1,\dots,\Delta x_N$, sampled at interval $\Delta t$, with covariance
matrix $\Sigma$. Let $\widehat{\Delta x}_k$ be the best linear prediction of
the $k$-th step from the $k-1$ steps before it (for Gaussian motion, its
conditional mean), and let
\begin{equation}
  v_k=\operatorname{Var}\bigl(\Delta x_k-\widehat{\Delta x}_k\bigr),
  \qquad k=1,\dots,N,
\end{equation}
be the variance of what remains unpredicted, the innovation. The first step
has nothing to be predicted from, so $v_1$ is the full step variance; each
later $v_k$ is smaller whenever the steps are correlated, and measures the
strength of the $k$-th fresh kick. A standard result of time-series analysis,
the prediction-error decomposition of the Gaussian
likelihood~\cite{brockwell1991time,hamilton1994time}, is that these variances
multiply to the determinant of the covariance,
\begin{equation}
  \det\Sigma=\prod_{k=1}^{N}v_k .
  \label{eq:innov}
\end{equation}
We therefore define the effective diffusion coefficient as the diffusivity of
the kicks,
\begin{equation}
  D_{\mathrm{eff}}=\frac{(\det\Sigma)^{1/N}}{2\Delta t}
  =\frac{1}{2\Delta t}\Bigl(\prod_{k=1}^{N}v_k\Bigr)^{1/N}.
  \label{eq:Deff_gen}
\end{equation}
For normal diffusion every step is a pure kick: the steps are uncorrelated,
each $v_k$ equals the step variance $2D\Delta t$, and $D_{\mathrm{eff}}=D$.
With memory, part of every step is predictable, and $D_{\mathrm{eff}}$ is
smaller than the apparent diffusivity
$D_{\mathrm{app}}=\langle|\Delta\mathbf r(\Delta t)|^2\rangle/(2d\Delta t)$
built from the raw step size. This holds for persistent and antipersistent
motion alike. $D_{\mathrm{eff}}$ is measured in $\mathrm L^2\,\mathrm T^{-1}$
for every process, so a change of time unit rescales all values by the same
factor and leaves the width and shape of their distribution unchanged.

The definition requires only that the step covariance exists, i.e.\ finite
second moments; the factorisation of Eq.~\eqref{eq:innov} holds for any
covariance matrix. For processes with stationary increments, such as FBM, the
covariance depends only on the separation between steps and can be estimated
by averaging along a trajectory. For processes that age, such as
continuous-time random walks or scaled Brownian motion, the covariance depends
on when the steps are taken, time averages along a trajectory no longer
estimate it, and $\Sigma_{jk}=\langle\Delta x_j\Delta x_k\rangle$ must instead
be averaged over many trajectories at the same times since a common origin;
$D_{\mathrm{eff}}$ then depends on that time. For non-Gaussian motion the kicks
are defined by the best linear prediction, and the interpretation holds at the
level of variances. In all cases $D_{\mathrm{eff}}$ measures how much
unpredictability the motion generates per unit time.

\subsubsection*{Fractional Brownian motion}
For FBM the step covariance is an amplitude multiplied by a correlation matrix
that depends only on the memory,
\begin{equation}
  \Sigma = 2D\,\Delta t^{2H}\,R(H),
  \label{eq:cov}
\end{equation}
where $R(H)$ is the correlation matrix of fractional Gaussian noise. The
effective diffusion coefficient then takes the closed form
\begin{equation}
  D_{\mathrm{eff}}
  = D\,\Delta t^{\,2H-1}\,\bigl[\det R(H)\bigr]^{1/N}
  = D_{\mathrm{app}}(\Delta t)\,f(H).
  \label{Seq:Deff}
\end{equation}
The factor $f(H)=[\det R(H)]^{1/N}$ is the average fraction of a step that the
preceding steps cannot predict: $1$ for Brownian motion, $0.96$ at $H=0.63$ and
$0.76$ at $H=0.78$ for $N=19$. For a strongly persistent cell, about a quarter
of each step is therefore inherited from its recent path, and $D_{\mathrm{eff}}$
counts only the remaining three quarters.

The same factor explains why $D_{\mathrm{eff}}$ removes the ridge of
Eq.~\eqref{eq:ridge}. How strongly the step correlations, and hence $f$, change
with $H$ determines how much information about $H$ comes from the correlations
rather than from the step size, and this places the pivot at
\begin{equation}
  \log\frac{\tau^*(H)}{\Delta t}=\frac{1}{2N}\,\frac{d\log\det R(H)}{dH}.
  \label{eq:taustar}
\end{equation}
For Brownian motion the steps are uncorrelated and the pivot sits at the frame
lag. For persistent motion stronger correlations pull it below $\Delta t$, and
for antipersistent motion they push it above: for $N=19$,
$\tau^*/\Delta t=1.46$, $1.00$, $0.49$ and $0.22$ at $H=0.3$, $0.5$, $0.7$ and
$0.8$. The ratio measured on simulated windows with $H=0.7$,
$0.51\pm0.01$, is the same from $\Delta t=0.02$ to $17$\,s and within $4\%$ of
the prediction. Writing $\log D_{\mathrm{eff}}=\log D+g(H)$, the slope of the
correction is exactly $g'(H)=2\log\tau^*(H)$: converting $D$ into
$D_{\mathrm{eff}}$ re-expresses the amplitude at the lag about which an error
in $H$ rotates the fitted MSD, so that errors in $H$ no longer propagate into
the amplitude. Reporting $D$ at one fixed reference lag instead removes the
ridge only for the single value of $H$ whose pivot coincides with that lag; the
apparent diffusivity $D_{\mathrm{app}}$, the choice $\Delta t$, works only for
Brownian motion. This separation of amplitude from memory is the finite-window
counterpart of a known result for stationary processes, in which the
innovation variance is asymptotically independent of the parameters that shape
the correlations~\cite{dahlhaus1989efficient,beran1994statistics}.

\subsubsection*{What $D_{\mathrm{eff}}$ is not}
$D_{\mathrm{eff}}$ describes the motion at the scale at which it is observed,
not the medium. It is not the prefactor of the mean squared displacement, which
at any lag is
\begin{equation}
  \langle|\Delta\mathbf r(t)|^2\rangle=2d\,D\,t^{2H}
  =2d\,\frac{D_{\mathrm{eff}}\,\Delta t}{f(H)}\left(\frac{t}{\Delta t}\right)^{2H}.
\end{equation}
Material properties defined through the MSD must therefore be computed from
$D$, or from $D_{\mathrm{eff}}$ after division by $f(H)$. For a passive probe
of radius $a$ in thermal equilibrium, for example, the generalized
Stokes--Einstein relation gives the creep compliance
$J(t)=(6\pi a/k_BT)\,D\,t^{2H}$, which reduces to $J=t/\eta$ in a Newtonian
fluid. Hemocytes are active, so no such relation applies to them: their
$D_{\mathrm{eff}}$ defines an effective temperature only through
$T_{\mathrm{eff}}=\gamma D_{\mathrm{eff}}/k_B$, and only relative changes are
meaningful while the friction $\gamma$ stays constant (main text). Finally, for
a self-similar process no diffusivity in $\mathrm L^2\,\mathrm T^{-1}$ can be
free of a time scale: $D_{\mathrm{eff}}$ is defined at the frame interval,
scales as $\Delta t^{2H-1}$, and depends weakly on the window length.
Comparisons with experiments at other frame rates must go back to $D$, which
reintroduces the uncertainty of $H$.

\subsubsection*{Measuring $D_{\mathrm{eff}}$}

For the hemocyte data, each track is divided into
overlapping windows of $20$ positions ($N=19$, $\Delta t=17.6$\,s). For each
window the network returns $H$ and $D$, and
\begin{equation}
  \log D_{\mathrm{eff}} = \log D + (2H-1)\log\Delta t
  + \tfrac1N\log\det R(H),
  \label{eq:Deff_hat}
\end{equation}
with $\log\det R(H)$ tabulated once on a fine grid of $H$. For example,
$H=0.7$ and $D=0.02\,\mu\mathrm m^2\,\mathrm s^{-1.4}$ give
$D_{\mathrm{eff}}\approx0.056\,\mu\mathrm m^2\,\mathrm s^{-1}$. Because the
network does not report how its errors in $H$ and $\log D$ covary, the
uncertainty of $\log D_{\mathrm{eff}}$ was measured on simulated windows at the
hemocyte frame interval (Methods). It is $0.080$--$0.092$\,dex over the whole
range of $H$, whereas the errors of $H$ and $\log D$ are correlated with
coefficient $-0.8$ to $-0.9$: the decoupling expected from the pivot argument
holds for the trained network. For time courses, the window values are combined
within each track and $176$\,s bin, and the median over tracks is reported with
a 95\% bootstrap interval. 
Since every window uses the same $\Delta t$ and $N$, values of
$D_{\mathrm{eff}}$ are directly comparable across developmental time and
between genotypes.

\section*{Supplementray figures}

A comparison of behaviour or diffusion coefficients with effective diffusion coefficients in  Fig.~\ref{fig:schedule}(A,B).  
\begin{figure}[t]
  \centering  \includegraphics[width=\columnwidth]{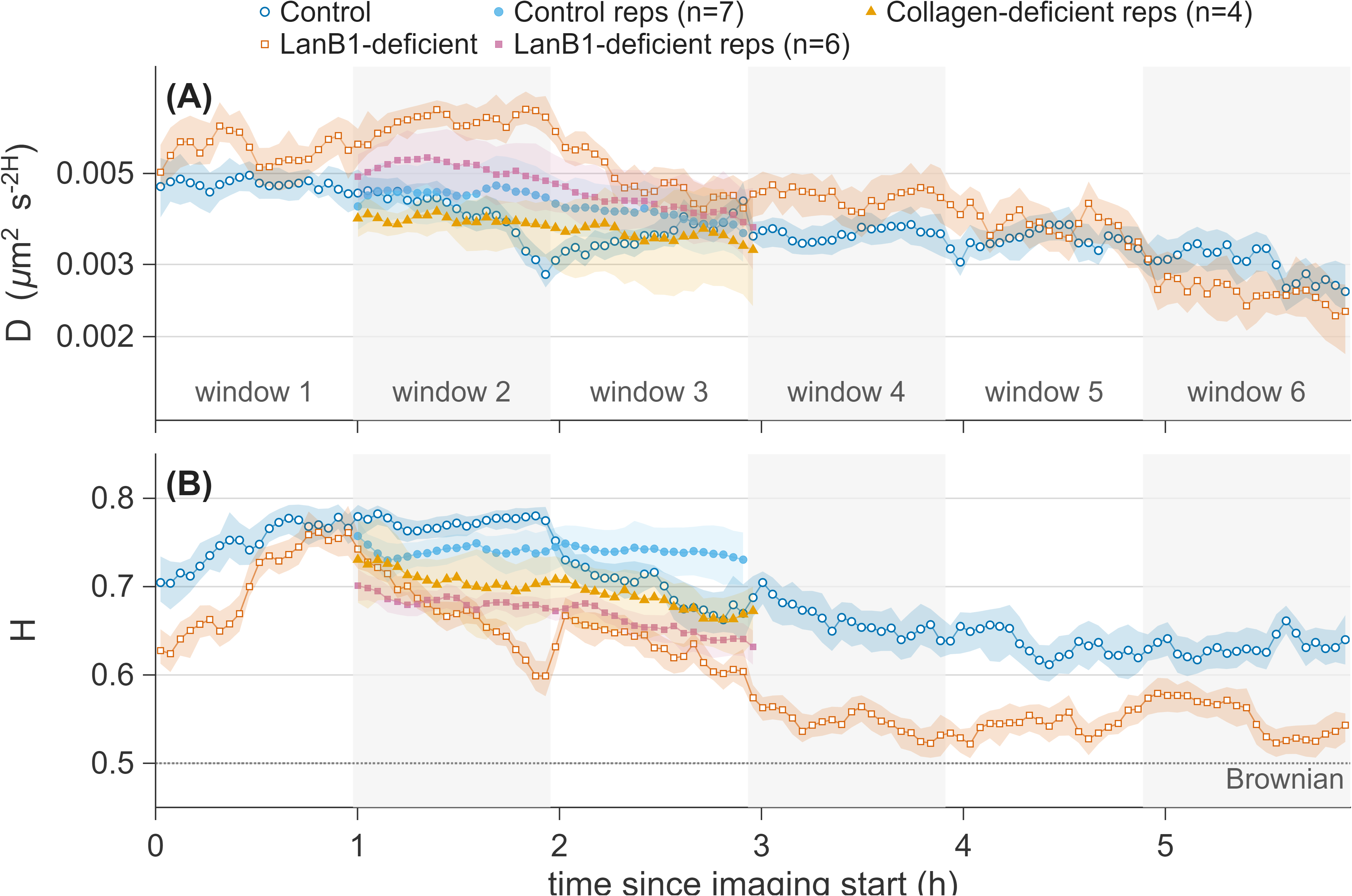}
  \caption{
  (A)~Diffusivity $D$ and
  (B)~Hurst exponent $H$, in 176\,s bins (10 frames) of the time since the
  start of imaging. Symbol shape encodes genotype: control (circles), \textit{LanB1}-deficient
  (squares) and collagen-deficient (triangles) embryos.
  Filled symbols, continuous six-movie recordings of control and \textit{LanB1}-deficient
  embryos; each point is the median over tracks with a 95\% bootstrap interval
  (shaded). Open symbols, mean over independent recordings: control ($n=7$)
  and \LanB{} ($n=4$), each spanning movies~1-2, and collagen mutant
  ($n=4$), two recordings spanning movies~1-2 and two spanning movies~2-3.
  Each open point is the mean of the per-recording medians, with a 95\%
  interval from a two-level bootstrap over recordings and over tracks within
  each recording. A bin is shown when at least 2 recordings contribute, each
  with at least 5 tracks.
  Dotted line in (B), Brownian motion, $H=0.5$. Estimates come from
  20-position windows ($\Delta t=17.6$\,s). Symbols are drawn every second
  bin.}
  \label{fig:schedule d}
\end{figure}

\end{document}